\documentclass{mn2e}
\usepackage{longtable}
\usepackage{graphicx}

\title[Type II Cepheid and RR Lyrae Luminosities] {The Luminosities 
and Distance Scales of
Type II Cepheid and RR Lyrae variables}
\author[Feast et al.]
{Michael W. Feast$^{1}$, Clifton D.Laney$^{2}$, Thomas D. Kinman$^{3}$,
Floor van Leeuwen$^{4}$,
\newauthor Patricia A. Whitelock$^{1,2,5}$\\
$^{1}$ Astronomy Department, University of Cape Town, 7701, Rondebosch,
South Africa.\\
(email: mwf@artemisia.ast.uct.ac.za)\\
$^{2}$ South African Astronomical Observatory, P.O. Box 9, 7935,
Observatory,South Africa.\\
$^{3}$ National Optical Astronomy Observatory, Tucson, 
P.O.Box 26732,  AZ85726, USA.\\
$^{4}$ Institute of Astronomy, Madingley Rd., Cambridge, England.\\
$^{5}$ National Astrophysics and Space Science Programme, Department
of Mathematics and Applied  Mathematics,\\
University of Cape Town, 7701,South Africa.\\}

\begin{document}
\maketitle
\begin{abstract}
Infrared and optical absolute magnitudes are derived for the type II
Cepheids $\kappa$ Pav and VY Pyx using revised Hipparcos parallaxes
and for $\kappa$ Pav, V553 Cen and SW Tau from pulsation parallaxes.
Revised Hipparcos and HST (Benedict et al.) parallaxes for RR Lyrae
agree satisfactorily and are combined in deriving absolute magnitudes.
Phase-corrected $J,H,K_{s}$ mags are given for 142 Hipparcos RR Lyraes
based on 2MASS observations. Pulsation and trigonometrical parallaxes
for classical Cepheids are compared to establish the best value
for the projection factor ($p$) used in pulsational analyses. 

The $M_{V}$ of RR Lyrae itself is $0.16\pm 0.12$ mag brighter than 
predicted
from a $M_{V}-[Fe/H]$ relation based  
RR Lyrae stars in the LMC at a modulus of $18.39
\pm 0.05$ as found from classical Cepheids. 
This is consistent with the prediction of Catalan \& Cort\'{e}s that it is
over luminous for its metallicity.
The $M_{K_{s}}$ results for 
the metal- and carbon-rich, Galactic disc stars,
V553 Cen and SW Tau, each with small internal errors ($\pm 0.08$ mag) have a mean
deviation of only 0.02 mag from the Period-Luminosity relation established
by Matsunaga et al. for type II Cepheids in globular clusters and with a
zero-point based on the same LMC scale. 
Comparing directly the luminosities of these two stars with published
data on Type II Cepheids in the LMC and in the Galactic Bulge leads
to an LMC modulus of $18.37 \pm 0.09$ and 
a distance to the Galactic Centre of $R_{0}=7.64 \pm 0.21$kpc.
The data for VY Pyx agree with
these results within the uncertainties set by its parallax.  Evidence is
presented that $\kappa$~Pav may have a close companion and possible
implications of this are discussed. If the pulsation parallax of this star
is incorporated in the analyses the distance scales just discussed will
be increased by $\sim 0.15 \pm 0.15$ mag.  
V553~Cen and SW~Tau show that at
optical wavelengths PL relations are wider for field stars than for those in
globular clusters. This is probably due to a narrower range of masses in the
latter case.

\end{abstract}
\section{Introduction}
 The RR Lyrae variables are known, primarily from studies of globular
clusters, to lie on or immediately above the Horizontal branch (HB) in an HR
diagram. Globular cluster studies also show a class of variable stars 
lying in an instability strip in an HR diagram
which extends approximately 3 mag above the HB.
%The fainter of these stars are thought
%to be evolving across this instability strip from the HB toward the
%AGB, whilst the brighter stars of the class are thought to be
%on blueward excursions from the AGB due to shell flashes 
%(Gingold 1976, 1985).
Variables with similar characteristics are also found in the general field,
both in the halo and disc. All these variables, both in clusters and the
field are classed together as ``type II Cepheids" (CephIIs).
These stars have been divided into three classes according to their periods.
Those of short period (roughly $P < 7$ days) are called BL Her stars, whilst
longer period ones (up to $P \sim 20$ days) are called W Vir stars. At even
longer periods, many of the CephIIs show characteristic alternations of deep
and shallow minima and are classed as RV Tau stars.  This subdivision of
CephIIs has not been universally adopted. Thus Sandage \& Tammann (2006)
review and summarize a system of classification based on light-curve
parameters that relate to their population characteristic and these
partially correlate with their metallicities. 
It should be noted that the ``population II Cepheids" with which
Sandage \& Tammann are primarily concerned are a subset of the
``type II Cepheids".
Maas et al. (2007) have shown
that the shorter period CephIIs in the general field differ from those of
longer period in their detailed chemical composition. The short period stars
are generally believed (Gingold 1976, 1985) to be evolving across the
instability strip from the HB towards the AGB. The longer period stars, on
the other hand, are believed to be on blueward excursions into the
instability strip from the AGB due to shell flashing.

In the present paper we discuss the luminosities of RR Lyrae and
CephIIs on the basis of the revised 
Hipparcos trigonometrical parallaxes (van Leeuwen 2007a,
see also van Leeuwen 2007b) and newly
derived pulsation parallaxes for three CephII variables.

%The ones of shorter period (roughly $ P < 7$ days) are called BL Her
%stars and are thought to be on their first crossing
%of the instability strip after they leave the HB. 
%Those of longer period 
%(up to $P \sim 20$ days)  are often called W Vir stars and are
%thought to be on shell-flashing loops. At even longer periods, many of
%the CepheidIIs show characteristic alternations of deep and shallow
%minima and are classed as RV Tau stars. They may be on flashing-loops
%or on their final blueward evolution at the end of the AGB phase.
%This subdivision of the CepheidIIs into three subclasses
%has not been universally adopted and other means of classification
%(e.g. by light curve shape) have been adopted (see the review by
%Sandage and Tammann 2006). Nevertheless the shorter period CepheidIIs
%in the general field are distinguished from those of longer period
%by their detailed chemical composition (Maas et al. 2007).

%An aim of the present paper is to investigate how far the recent
%revision of the Hipparcos parallaxes (van Leeuwen 2007) helps
%establish luminosities of the CepheidIIs and RR Lyrae variables and 
%to study the consequences of these
%results. We also present the first pulsation
%parallaxes (Baade-Wesselink method) of 3 CephII 
%variables and compare them with the results derived from the 
%parallax work.

%\section{Data}

\section{Period-Luminosity and Metallicity-Luminosity Relations}
 Here we present 
various relations which are required
in the interpretation of our data.
\subsection{Relationships for RR Lyrae variables}
It has long been thought that the luminosities of RR Lyrae variables
can be expressed in the form:
\begin{equation}
 M_{V} = a [Fe/H] + b
\end{equation}
However, the values of $a$ and $b$ have been much disputed, as has 
the question of the linearity of the equation.
In the following we adopt:
\begin{equation}
M_{V} = 0.214 [Fe/H] + (19.39 -Mod(LMC)).
\end{equation}
This is based on RR Lyraes in the LMC (Gratton et al. 2004).  Adopting an
LMC modulus of 18.39\footnote{This includes a correction for metallicity
effects based on Marci et al. 2006.} as derived from classical Cepheids
(Benedict et al. 2007, van Leeuwen et al. 2007), the constant term
becomes:\\
$b=+1.0$.\\
%This modulus is derived using (classical) Cepheids and its main
%uncertainty is the correction  for the difference in metallicity
%of Galactic and LMC Cepheids. Without such a correction the modulus
%of the LMC would be 18.50 and the constant term, $b$, would be 0.89.

The LMC RR Lyraes, on which this relation is based, cover a range in
[Fe/H] from $\sim -0.8$ to $-2.2$, but are mainly
concentrated between $-1.3$ and $-1.8$.  There is evidence, however,
that the slope of the relation is not universal. Clementini et al. (2005)
find that in the Sculptor dwarf spheroidal,
over roughly the same metallicity range, the slope is  
$0.092 \pm 0.027$ compared with the LMC $0.214 \pm 0.047$ and they
suggest that the Sculptor RR Lyraes are on average more evolved
than those in the LMC.

That there is a period-luminosity relation for RR Lyraes in the $K$ band
(PL($K$)), possibly independent, or nearly independent
of metallicity, goes back at least  to the work of Longmore et al. (1986) on
globular clusters. The most recent version of such a relation was given by
Sollima et al. (2006) again based on globular clusters. The relative
distances of the clusters came from main-sequence fitting and the zero-point
of their final relation was set by a trigonometrical parallax of RR Lyrae
itself (Benedict et al. 2002). They found:
\begin{eqnarray}
M_{K_{s}} = -2.38 (\pm 0.04) \log P + 0.08 (\pm 0.11) [Fe/H] - \nonumber\\
1.05 (\pm 0.13),
\end{eqnarray}
where $K_{s}$ is the $K_{s}$ magnitude in the 2MASS system.
The term in [Fe/H] is small and not statistically significant.
\subsection{Relationships for type II Cepheids}
 In the past various PL relations for CephIIs at visual wavelengths have
been suggested based primarily on globular cluster work. More recently, it
was shown from globular cluster data that a well defined PL($K_{s}$)
relation, with small scatter, applied (Matsunaga et al. 2006). The globular
cluster distances were determined from a relation 
for horizontal branch stars similar to
eq. 2 and we may write the Matsunaga CephII relation as:
%but with a constant term of
%0.89, leading (Gratton et al. 2003) to an LMC modulus of $18.5 \pm 0.09$.
%Within the errors this is not inconsistent with the LMC Cepheid modulus
%mentioned above. However, to make a consistent comparison of different
%distance scales possible we have adjusted the result of Matsunaga et al.
%by +0.11mag. The relation for CepheidIIs in globular clusters thus becomes:
\begin{equation}
M_{K_{s}} = -2.41 (\pm 0.05) \log P + c,
\end{equation}
where $c = 17.39 - Mod(LMC)$\\
and $c = -1.0$ for $Mod(LMC) = 18.39$ as above.\\
The (internal) standard error of the constant term is $\pm 0.02$
at the mean $\log P$ (1.120).  
Matsunaga et al. pointed out that
RR Lyraes in clusters lay on an extrapolation of this relation to
shorter periods. The subsequent work of Sollima et al. (2006) 
confirms
this (compare eqs.  3 and 4). Matsunaga et al. examined their
data for a metallicity effect on the PL($K$) zero-point and found a term,
$-0.10 \pm 0.06$. This is clearly not significant and is of
opposite sign to the metallicity term in the RR Lyrae relation (eq. 3) which
is also not significant. This suggests that a combined RR Lyrae/CephII
PL($K$) is virtually metal independent in globular clusters. Some caution is
necessary in accepting this, however, since there are only four CephIIs in
the Matsunaga sample with $[Fe/H] > -1.0$ and these all have periods greater
than 10 days\footnote{But see the discussion of the field variables below.}.

In addition to the above the following three PL relations at
optical wavelengths 
will be required later. They are based on CephIIs in NGC\,6441
and 6388 and are taken directly
from Pritzl et al. (2003). 
\begin{equation}
M_{V} = -1.64(\pm 0.05) \log P + 0.05 (\pm 0.05),  
\end{equation}

\begin{equation}
M_{B} = -1.23(\pm 0.09) \log P + 0.31 (\pm 0.09),
\end{equation}

\begin{equation}
M_{I} = -2.03(\pm 0.03) \log P - 0.36 (\pm 0.01).  
\end{equation}
   
\section{The RR Lyrae Variables}
\subsection{Data}
Table~\ref{142rr} lists the data for 142 RR Lyrae variables. 
%TABLE 1
\begin{center}
\onecolumn
\begin{longtable}{rrrcrrrrcccc}
\caption[Basic data used in the analysis.]{Basic data used in the
analysis.}\label{142rr} \\
\hline
\multicolumn{1}{c}{Hipparcos} & \multicolumn{1}{c}{name} & \multicolumn{1}{c}{$\pi$} & $\Delta \pi$
& \multicolumn{1}{c}{$V$} & \multicolumn{1}{c}{$J$} & \multicolumn{1}{c}{$H$}& 
\multicolumn{1}{c}{$K_{s}$} & P & [Fe/H] & E$_{(B-V)}$ & typ\\
& &  \multicolumn{2}{c}{(mas)}&  \multicolumn{4}{c}{(mag)} & (day)& &
(mag)\\
\hline
\endfirsthead

\hline
\multicolumn{1}{c}{Hipparcos} & \multicolumn{1}{c}{name} & \multicolumn{1}{c}{$\pi$} & $\Delta \pi$
& \multicolumn{1}{c}{$V$} & \multicolumn{1}{c}{$J$} & \multicolumn{1}{c}{$H$}& 
\multicolumn{1}{c}{$K_s$} & P & [Fe/H] & E$_{(B-V)}$ & typ\\
& &  \multicolumn{2}{c}{(mas)}&  \multicolumn{4}{c}{(mag)} & (day)& &
(mag)\\
\hline
\endhead
%This is the footer for all pages except the last page of the table...
  \multicolumn{12}{l}{{Continued on Next Page\ldots}} \\
\endfoot    

%This is the footer for the last page of the table...
  \\ \hline 
\endlastfoot

   226 & RU Scl     &  0.99 & 1.96 & 10.220 &  9.474 &  9.294 &  9.229 & 0.493347 & --1.27 & 0.018 &   \\
   320 & UU Cet     &  1.59 & 5.73 & 12.080 & 11.137 & 10.863 & 10.837 & 0.606080 & --1.28 & 0.021 &   \\
  1878 & SW And     & --0.01 & 1.84 &  9.710 &  8.809 &  8.578 &  8.505 & 0.442262 & --0.24 & 0.038 &   \\
  2655 & RX Cet     &  3.24 & 4.74 & 11.440 & 10.606 & 10.378 & 10.319 & 0.573685 & --1.28 & 0.025 &   \\
  4541 & W Tuc      &  5.37 & 2.41 & 11.410 & 10.594 & 10.373 & 10.344 & 0.642260 & --1.57 & 0.021 &   \\
  4725 & RU Cet     &  7.14 & 4.62 & 11.680 & 10.597 & 10.487 & 10.465 & 0.586267 & --1.66 & 0.023 &   \\
  5803 & RU Psc     &  1.30 & 2.08 & 10.190 &  9.347 &  9.162 &  9.117 & 0.390333 & --1.75 & 0.043 & c \\
  6029 & XX And     & --0.79 & 2.50 & 10.680 &  9.727 &  9.488 &  9.409 & 0.722755 & --1.94 & 0.039 &   \\
  6094 & VW Scl     &  2.34 & 2.79 & 11.030 & 10.418 & 10.193 & 10.136 & 0.510913 & --0.84 & 0.016 &   \\
  6115 & AM Tuc     & --1.93 & 2.28 & 11.670 & 10.865 & 10.617 & 10.563 & 0.405769 & --1.49 & 0.023 & c \\
  7149 & RR Cet     &  0.48 & 1.85 &  9.730 &  8.829 &  8.623 &  8.520 & 0.553030 & --1.45 & 0.022 &   \\
  7398 & VX Scl     &  3.71 & 3.64 & 12.020 & 11.094 & 10.894 & 10.853 & 0.637058 & --2.25 & 0.014 &   \\
  8163 & SV Scl     &  5.50 & 2.37 & 11.380 & 10.718 & 10.596 & 10.543 & 0.377380 & --1.77 & 0.014 & c \\
  8939 & CI And     &  0.77 & 5.87 & 12.280 & 11.182 & 11.018 & 11.185 & 0.484728 & --0.69 & 0.062 &   \\
  9932 & SS For     &  3.57 & 1.98 & 10.190 &  9.546 &  9.305 &  9.246 & 0.495424 & --0.94 & 0.014 &   \\
 10491 & RV Cet     &  2.16 & 2.70 & 10.920 &  9.903 &  9.580 &  9.520 & 0.623350 & --1.60 & 0.024 &   \\
 11517 & RZ Cet     & --0.04 & 4.92 & 11.850 & 11.031 & 10.787 & 10.737 & 0.510606 & --1.36 & 0.029 &   \\
 12199 & CS Eri     &  2.70 & 1.10 &  9.000 &  8.144 &  8.014 &  7.973 & 0.311332 & --1.41 & 0.018 & c \\
 14601 & X Ari      &  0.99 & 1.90 &  9.570 &  8.365 &  8.042 &  7.941 & 0.651154 & --2.43 & 0.180 &   \\
 14856 & SV Eri     &  3.18 & 2.53 &  9.960 &  8.958 &  8.710 &  8.642 & 0.713865 & --1.70 & 0.085 &   \\
 16321 & SX For     & --5.39 & 2.38 & 11.120 & 10.035 &  9.847 &  9.772 & 0.605342 & --1.66 & 0.012 &   \\
 19993 & AR Per     & --1.32 & 2.02 & 10.510 &  9.012 &  8.710 &  8.642 & 0.425551 & --0.30 & 0.108 &   \\
 22442 & RX Eri     &  1.31 & 1.70 &  9.690 &  8.737 &  8.485 &  8.429 & 0.587246 & --1.33 & 0.058 &   \\
 22466 & U Pic      &  3.21 & 2.21 & 11.380 & 10.689 & 10.464 & 10.381 & 0.440373 & --0.72 & 0.009 &   \\
 22750 & BB Eri     &  5.44 & 3.58 & 11.520 & 10.321 & 10.147 & 10.110 & 0.569909 & --1.32 & 0.048 &   \\
 22952 & U Lep      &  2.32 & 2.97 & 10.570 &  9.814 &  9.565 &  9.542 & 0.581479 & --1.78 & 0.027 &   \\
 24471 & RY Col     &  3.35 & 1.79 & 10.900 & 10.254 &  9.732 &  9.699 & 0.478832 & --0.91 & 0.026 &   \\
 29528 & RX Col     & --4.02 & 5.53 & 12.720 & 11.634 & 11.393 & 11.313 & 0.593780 & --1.70 & 0.082 &   \\
 34743 & TZ Aur     & --3.70 & 6.39 & 11.910 & 10.975 & 10.771 & 10.731 & 0.391676 & --0.79 & 0.037 &   \\
 35281 & AA CMi     &  1.40 & 5.22 & 11.570 & 10.570 & 10.384 & 10.281 & 0.476327 & --0.15 & 0.011 &   \\
 35584 & HH Pup     &  2.39 & 2.53 & 11.290 & 10.248 & 10.044 &  9.975 & 0.390748 & --0.50 & 0.158 &   \\
 35667 & RR Gem     &  0.43 & 3.24 & 11.380 & 10.566 & 10.306 & 10.275 & 0.397292 & --0.29 & 0.054 &   \\
 37779 & HK Pup     & --2.90 & 3.60 & 11.370 & 10.240 & 10.010 &  9.915 & 0.734229 & --1.11 & 0.160 &   \\
 37805 & TW Lyn     & --6.62 & 8.60 & 12.000 & 11.075 & 10.854 & 10.778 & 0.481862 & --0.66 & 0.051 &   \\
 38561 & SZ Gem     &  6.04 & 4.19 & 11.750 & 11.072 & 10.798 & 10.748 & 0.501143 & --1.46 & 0.013 &   \\
 39009 & UY Cam     &  0.19 & 1.99 & 11.530 & 11.002 & 10.872 & 10.859 & 0.267044 & --1.33 & 0.022 & c \\
 39849 & XX Pup     & --0.15 & 3.81 & 11.250 & 10.321 & 10.118 & 10.084 & 0.517203 & --1.33 & 0.068 &   \\
 40186 & DD Hya     & --5.41 & 5.88 & 12.220 & 11.457 & 11.241 & 11.228 & 0.501771 & --0.97 & 0.013 &   \\
 41936 & TT Cnc     &  2.42 & 5.55 & 11.350 & 10.330 & 10.047 &  9.968 & 0.563430 & --1.57 & 0.043 &   \\
 44428 & TT Lyn     & --1.48 & 1.75 &  9.860 &  8.908 &  8.655 &  8.611 & 0.597429 & --1.56 & 0.017 &   \\
 45709 & RW Cnc     &  1.05 & 4.98 & 11.850 & 10.677 & 10.561 & 10.530 & 0.547224 & --1.67 & 0.020 &   \\
 48503 & T Sex      &  2.24 & 1.56 & 10.040 &  9.438 &  9.268 &  9.200 & 0.324706 & --1.34 & 0.044 & c \\
 49628 & RR Leo     &  5.01 & 3.16 & 10.730 & 10.021 &  9.778 &  9.730 & 0.452392 & --1.60 & 0.037 &   \\
 50073 & WZ Hya     &  4.50 & 5.17 & 10.900 &  9.945 &  9.669 &  9.610 & 0.537713 & --1.39 & 0.075 &   \\
 50289 & WY Ant     & --0.27 & 2.51 & 10.870 &  9.970 &  9.744 &  9.674 & 0.574341 & --1.48 & 0.059 &   \\
 53213 & AF Vel     &  0.57 & 3.19 & 11.440 & 10.354 & 10.079 & 10.040 & 0.527414 & --1.49 & 0.250 &   \\
 55825 & W Crt      & --1.95 & 3.43 & 11.540 & 10.774 & 10.590 & 10.539 & 0.412015 & --0.54 & 0.040 &   \\
 56088 & TU UMa     &  0.56 & 1.68 &  9.820 &  8.919 &  8.740 &  8.660 & 0.557658 & --1.51 & 0.022 &   \\
 56350 & AX Leo     & --3.10 & 7.00 & 12.260 & 11.302 & 11.048 & 10.951 & 0.726845 & --1.72 & 0.033 &   \\
 56409 & SS Leo     &  2.50 & 4.01 & 11.030 & 10.259 & 10.008 &  9.943 & 0.626335 & --1.79 & 0.018 &   \\
 56734 & SU Dra     &  1.27 & 1.53 &  9.780 &  8.898 &  8.676 &  8.619 & 0.660418 & --1.80 & 0.010 &   \\
 56742 & BX Leo     &  7.73 & 6.17 & 11.610 & 10.889 & 10.743 & 10.709 & 0.362757 & --1.28 & 0.023 & c \\
 56785 & ST Leo     & --0.45 & 3.47 & 11.490 & 10.690 & 10.480 & 10.446 & 0.477990 & --1.17 & 0.038 &   \\
 57625 & X Crt      & --3.98 & 4.50 & 11.480 & 10.482 & 10.213 & 10.148 & 0.732842 & --2.00 & 0.027 &   \\
 58907 & IK Hya     &  1.39 & 1.62 & 10.110 &  9.144 &  8.863 &  8.760 & 0.650371 & --1.24 & 0.061 &   \\
 59208 & UU Vir     &  2.24 & 2.91 & 10.560 &  9.596 &  9.436 &  9.414 & 0.475597 & --0.87 & 0.018 &   \\
 59411 & AB UMa     &  0.12 & 1.94 & 10.940 &  9.934 &  9.678 &  9.623 & 0.599593 & --0.49 & 0.022 &   \\
 59946 & SW Dra     &  2.24 & 1.42 & 10.480 &  9.594 &  9.362 &  9.319 & 0.569671 & --1.12 & 0.014 &   \\
 61029 & UZ CVn     &  6.50 & 7.59 & 12.120 & 11.219 & 10.941 & 10.885 & 0.697791 & --1.89 & 0.019 &   \\
 61031 & SV Hya     &  3.79 & 2.16 & 10.530 &  9.673 &  9.455 &  9.366 & 0.478542 & --1.50 & 0.080 &   \\
 61225 & S Com      &  5.16 & 3.66 & 11.630 & 10.823 & 10.678 & 10.619 & 0.586585 & --1.91 & 0.019 &   \\
 61809 & U Com      &  7.40 & 4.05 & 11.740 & 11.186 & 10.984 & 10.987 & 0.292736 & --1.25 & 0.014 & c \\
 63054 & AT Vir     &  1.32 & 3.03 & 11.340 & 10.547 & 10.363 & 10.332 & 0.525785 & --1.60 & 0.030 &   \\
 64875 & ST Com     & --3.68 & 3.55 & 11.460 & 10.461 & 10.258 & 10.186 & 0.598927 & --1.10 & 0.024 &   \\
 65063 & AV Vir     &  2.22 & 4.73 & 11.820 & 10.853 & 10.615 & 10.566 & 0.656910 & --1.25 & 0.028 &   \\
 65344 & AM Vir     & --1.79 & 3.17 & 11.520 & 10.509 & 10.253 & 10.199 & 0.615063 & --1.37 & 0.067 &   \\
 65445 & AU Vir     &  0.06 & 4.99 & 11.590 & 11.085 & 10.918 & 10.847 & 0.339616 & --1.50 & 0.028 & c \\
 65547 & SX UMa     &  1.90 & 1.81 & 10.840 & 10.288 & 10.135 & 10.071 & 0.307139 & --1.81 & 0.010 & c \\
 66122 & RV UMa     & --0.30 & 1.85 & 10.770 & 10.058 &  9.854 &  9.831 & 0.468069 & --1.20 & 0.018 &   \\
 67087 & RZ CVn     & --2.03 & 2.99 & 11.570 & 10.733 & 10.518 & 10.478 & 0.567403 & --1.84 & 0.014 &   \\
 67227 & RV Oct     &  1.75 & 2.17 & 10.980 &  9.879 &  9.614 &  9.526 & 0.571169 & --1.71 & 0.180 &   \\
 67354 & SS CVn     &  2.14 & 3.83 & 11.840 & 11.185 & 10.951 & 10.936 & 0.478510 & --1.37 & 0.006 &   \\
 67976 & V499 Cen   & --0.01 & 2.97 & 11.120 & 10.225 &  9.926 &  9.922 & 0.521205 & --1.43 & 0.085 &   \\
 68188 & ST CVn     & --1.28 & 4.11 & 11.370 & 10.626 & 10.459 & 10.449 & 0.329065 & --1.07 & 0.012 & c \\
 68292 & UY Boo     &  1.45 & 3.00 & 10.940 &  9.981 &  9.755 &  9.723 & 0.650889 & --2.56 & 0.033 &   \\
 68908 & W CVn      &  2.95 & 2.42 & 10.550 &  9.667 &  9.454 &  9.371 & 0.551753 & --1.22 & 0.005 &   \\
 69759 & TV Boo     & --0.05 & 2.09 & 10.970 & 10.373 & 10.282 & 10.248 & 0.312557 & --2.44 & 0.010 & c \\
 70702 & ST Vir     & --5.10 & 5.66 & 11.520 & 10.914 & 10.748 & 10.671 & 0.410806 & --0.67 & 0.039 &   \\
 70751 & AF Vir     & --9.08 & 5.23 & 11.800 & 10.939 & 10.769 & 10.684 & 0.483735 & --1.33 & 0.023 &   \\
 71186 & RS Boo     &  1.62 & 1.91 & 10.370 &  9.744 &  9.559 &  9.507 & 0.377339 & --0.36 & 0.012 &   \\
 72115 & TW Boo     & --2.23 & 2.28 & 11.290 & 10.407 & 10.192 & 10.170 & 0.532277 & --1.46 & 0.013 &   \\
 72342 & AE Boo     &  0.33 & 2.00 & 10.650 &  9.974 &  9.819 &  9.762 & 0.314893 & --1.39 & 0.023 & c \\
 72444 & TY Aps     &  1.78 & 3.07 & 11.850 & 10.819 & 10.532 & 10.456 & 0.501695 & --0.95 & 0.169 &   \\
 72691 & BT Dra     & --1.26 & 2.08 & 11.640 & 10.735 & 10.478 & 10.397 & 0.588673 & --1.75 & 0.010 &   \\
 72721 & XZ Aps     & --4.19 & 5.48 & 12.380 & 11.284 & 11.006 & 10.923 & 0.587275 & --1.06 & 0.135 &   \\
 74556 & AP Ser     & --0.16 & 4.32 & 11.110 & 10.462 & 10.305 & 10.268 & 0.340805 & --1.58 & 0.042 & c \\
 75225 & TV CrB     &  1.89 & 5.75 & 11.870 & 11.037 & 10.814 & 10.774 & 0.584629 & --2.33 & 0.033 &   \\
 75234 & FW Lup     &  1.58 & 1.18 &  9.060 &  7.995 &  7.836 &  7.671 & 0.484169 & --0.20 & 0.077 &   \\
 75942 & ST Boo     & --0.13 & 1.80 & 11.010 & 10.185 &  9.981 &  9.930 & 0.622286 & --1.76 & 0.021 &   \\
 75982 & VY Ser     & --0.77 & 1.99 & 10.130 &  9.205 &  8.944 &  8.826 & 0.714101 & --1.79 & 0.040 &   \\
 76313 & CG Lib     & --0.50 & 5.67 & 11.550 & 10.437 & 10.208 & 10.125 & 0.306787 & --1.19 & 0.297 & c \\
 77663 & VY Lib     & --1.84 & 4.04 & 11.730 & 10.480 & 10.174 & 10.070 & 0.533941 & --1.34 & 0.192 &   \\
 77830 & AN Ser     & --4.47 & 4.79 & 10.940 & 10.096 &  9.898 &  9.842 & 0.522069 & --0.07 & 0.040 &   \\
 77997 & AT Ser     &  0.18 & 5.30 & 11.480 & 10.533 & 10.248 & 10.214 & 0.746570 & --2.03 & 0.037 &   \\
 78417 & AR Her     &  2.08 & 3.25 & 11.240 & 10.605 & 10.413 & 10.391 & 0.469981 & --1.30 & 0.013 &   \\
 79974 & RV CrB     &  3.77 & 3.21 & 11.410 & 10.555 & 10.418 & 10.336 & 0.331593 & --1.69 & 0.039 & c \\
 80402 & V445 Oph   &  5.60 & 5.33 & 11.050 &  9.649 &  9.401 &  9.262 & 0.397023 & --0.19 & 0.287 &   \\
 80853 & VX Her     & --0.78 & 2.65 & 10.690 &  9.848 &  9.651 &  9.590 & 0.455362 & --1.58 & 0.044 &   \\
 80990 & UV Oct     &  2.32 & 1.12 &  9.500 &  8.592 &  8.362 &  8.297 & 0.542587 & --1.74 & 0.091 &   \\
 81238 & RW Dra     &  1.38 & 2.44 & 11.710 & 10.779 & 10.596 & 10.622 & 0.442909 & --1.55 & 0.011 &   \\
 83244 & RW TrA     &  5.74 & 3.19 & 11.400 & 10.375 & 10.111 & 10.059 & 0.374039 & --0.13 & 0.105 &   \\
 84233 & VZ Her     &  3.49 & 2.12 & 11.480 & 10.746 & 10.590 & 10.496 & 0.440331 & --1.02 & 0.027 &   \\
 87681 & TW Her     & --3.36 & 2.22 & 11.280 & 10.528 & 10.322 & 10.239 & 0.399599 & --0.69 & 0.042 &   \\
 87804 & WY Pav     &  1.08 & 6.99 & 12.180 & 10.836 & 10.647 & 10.553 & 0.588573 & --0.98 & 0.126 &   \\
 88064 & S Ara      & --2.11 & 3.31 & 10.780 &  9.867 &  9.601 &  9.560 & 0.451879 & --0.71 & 0.124 &   \\
 88402 & MS Ara     &  8.81 & 5.20 & 12.070 & 11.036 & 10.763 & 10.664 & 0.524982 & --1.48 & 0.146 &   \\
 89326 & V675 Sgr   & --1.28 & 2.75 & 10.330 &  9.313 &  9.053 &  9.003 & 0.642280 & --2.28 & 0.130 &   \\
 89372 & BC Dra     &  1.51 & 1.99 & 11.600 & 10.435 & 10.172 & 10.096 & 0.719590 & --2.00 & 0.068 &   \\
 89450 & V455 Oph   & --1.47 & 6.69 & 12.360 & 11.395 & 11.160 & 11.088 & 0.453882 & --1.07 & 0.144 &   \\
 90053 & IO Lyr     & --0.84 & 2.95 & 11.850 & 10.841 & 10.591 & 10.538 & 0.577121 & --1.14 & 0.074 &   \\
 91634 & CN Lyr     & --3.91 & 2.52 & 11.480 & 10.282 & 10.055 &  9.919 & 0.411383 & --0.58 & 0.178 &   \\
 92244 & V413 CrA   & --1.75 & 3.26 & 10.600 &  9.497 &  9.248 &  9.148 & 0.589343 & --1.26 & 0.075 &   \\
 93476 & MT Tel     &  1.17 & 1.46 &  8.980 &  8.323 &  8.176 &  8.076 & 0.316900 & --1.85 & 0.038 & c \\
 94134 & XZ Dra     &  2.28 & 1.20 & 10.250 &  9.398 &  9.221 &  9.148 & 0.476497 & --0.79 & 0.062 &   \\
 94869 & BK Dra     &  0.67 & 1.52 & 11.190 & 10.336 & 10.124 & 10.071 & 0.592076 & --1.95 & 0.052 &   \\
 95497 & RR Lyr     &  3.79 & 0.19 &  7.760 &  6.759 &  6.546 &  6.489 & 0.566839 & --1.39 & 0.030 &   \\
 95702 & BN Vul     &  3.56 & 3.08 & 11.020 &  9.138 &  8.793 &  8.677 & 0.594138 & --1.61 & 0.173 &   \\
 96101 & V440 Sgr   & --0.09 & 3.43 & 10.340 &  9.402 &  9.153 &  9.082 & 0.477479 & --1.40 & 0.085 &   \\
 96112 & XZ Cyg     &  1.83 & 1.01 &  9.680 &  8.990 &  8.793 &  8.722 & 0.466610 & --1.44 & 0.096 &   \\
 96581 & BN Pav     &  6.43 & 6.05 & 12.600 & 11.593 & 11.344 & 11.279 & 0.567117 & --1.32 & 0.073 &   \\
 98265 & BP Pav     &  3.50 & 6.34 & 12.540 & 11.648 & 11.386 & 11.366 & 0.527128 & --1.48 & 0.059 &   \\
101356 & V341 Aql   & --4.86 & 5.62 & 10.850 &  9.886 &  9.687 &  9.606 & 0.578017 & --1.22 & 0.086 &   \\
102593 & DX Del     &  0.40 & 1.94 &  9.940 &  9.048 &  8.746 &  8.685 & 0.472619 & --0.39 & 0.092 &   \\
103364 & UY Cyg     &  2.55 & 2.91 & 11.110 & 10.060 &  9.805 &  9.777 & 0.560714 & --0.80 & 0.129 &   \\
103755 & RV Cap     &  0.85 & 3.82 & 11.040 &  9.703 &  9.717 &  9.753 & 0.447698 & --1.61 & 0.041 &   \\
104613 & V Ind      &  1.09 & 2.06 &  9.960 &  9.274 &  9.028 &  8.985 & 0.479604 & --1.50 & 0.043 &   \\
104930 & SW Aqr     & --3.93 & 4.09 & 11.180 & 10.413 & 10.142 & 10.057 & 0.459299 & --1.63 & 0.076 &   \\
105026 & Z Mic      &  0.69 & 3.53 & 11.650 & 10.478 & 10.179 & 10.112 & 0.586925 & --1.10 & 0.094 &   \\
105285 & YZ Cap     &  4.62 & 2.78 & 11.300 & 10.532 & 10.437 & 10.429 & 0.273461 & --1.06 & 0.063 & c \\
106645 & SX Aqr     &  2.42 & 3.58 & 11.780 & 10.973 & 10.689 & 10.639 & 0.535712 & --1.87 & 0.048 &   \\
106649 & RY Oct     & --1.87 & 4.88 & 12.060 & 11.118 & 10.917 & 10.859 & 0.563475 & --1.83 & 0.113 &   \\
107078 & CG Peg     &  3.16 & 2.49 & 11.180 & 10.216 & 10.007 &  9.970 & 0.467133 & --0.50 & 0.074 &   \\
107935 & AV Peg     &  2.88 & 2.44 & 10.500 &  9.609 &  9.406 &  9.346 & 0.390378 & --0.08 & 0.067 &   \\
108057 & SS Oct     &  9.09 & 3.32 & 11.910 & 10.041 &  9.835 &  9.752 & 0.621852 & --1.60 & 0.285 &   \\
108839 & BV Aqr     &  7.24 & 4.15 & 10.900 & 10.228 & 10.017 & 10.075 & 0.363653 & --1.42 & 0.034 &   \\
111839 & RZ Cep     &  0.60 & 1.48 &  9.470 &  8.168 &  7.959 &  7.883 & 0.308688 & --1.77 & 0.078 & c \\
112994 & BH Peg     & --0.72 & 2.38 & 10.460 &  9.385 &  9.114 &  9.067 & 0.640991 & --1.22 & 0.077 &   \\
115135 & DN Aqr     & --1.08 & 2.82 & 11.200 & 10.158 &  9.934 &  9.900 & 0.633757 & --1.66 & 0.025 &   \\
115870 & RV Phe     &  1.75 & 4.71 & 11.940 & 11.106 & 10.828 & 10.768 & 0.596416 & --1.69 & 0.007 &   \\
116664 & BR Aqr     &  0.71 & 3.48 & 11.420 & 10.648 & 10.421 & 10.370 & 0.481872 & --0.74 & 0.027 &   \\
116942 & VZ Peg     &  4.89 & 3.75 & 11.900 & 11.219 & 11.059 & 11.010 & 0.306493 & --1.80 & 0.045 & c \\
116958 & AT And     & --2.25 & 1.85 & 10.710 &  9.478 &  9.181 &  9.087 & 0.616917 & --1.18 & 0.110 &   \\

\end{longtable}
\end{center}
\twocolumn  
The stars are those listed by Fernley et al. (1998) and we have generally
adopted their $V$ magnitudes and [Fe/H] values. The parallaxes and their
standard errors are from the revised Hipparcos catalogue (van Leeuwen 2007). 
Details regarding the formation of the table, particularly the derivation of
mean $JHK_{s}$ values from the single 2MASS values, are given in Appendix
B\footnote{Since our analysis of the RR Lyrae data was completed, Sollima et
al. (2007) have published mean $J,H,K_{s}$ data for RR Lyrae itself. They
measured against 2MASS stars as standards and found: $6.74 \pm 0.02$, $6.60
\pm 0.03$ and $6.50 \pm 0.02$. The values we derived (Table~1) are 6.76,
6.55 and 6.49. The Sollima et al. results provide a useful confirmation of
our procedure. Since their value of $K_{s}$ is negligibly different from our
value we have kept our value in the following.}. DH Peg, which is in the
Fernley et al. list, has been omitted because its status is doubtful. It may
be a dwarf Cepheid (Fernley et al. 1990).  There are a number of other stars
which are listed as RR Lyrae stars in the Hipparcos catalogue. In some cases
this classification is incorrect or doubtful. For instance DX Cet is
actually a $\delta$~Sct star (Kiss et al. 1999).  This star is, in fact, of
special interest as having a parallax with a small percentage error and
falling on the PL relation for fundamental mode $\delta$ Sct pulsators (van
Leeuwen 2007).  A discussion of stars whose classification as RR Lyrae type
is probably incorrect or uncertain will be given elsewhere (Kinman, in
preparation). The parallaxes and magnitudes of the very few Hipparcos stars
which are probably RR Lyraes and were not in the Fernley list are such that
they would make no significant contribution to the results given in this
paper. It seemed better therefore to omit them and thus, for instance, have
the homogeneous set of [Fe/H] results given by Fernley et al. The
reddenings, $E(B-V)$, listed are the means of the two values discussed in
section 3.2. These two values agree closely, the maximum difference (0.06
mag) being that for BN Vul, a star at low galactic latitude. For RZ Cep,
which is also close to the plane, the difference is 0.03 mag. All other
stars show smaller differences.

We assume in the following that,\\
$A_{V} = 3.06E(B-V)$\\
and with data on the 2MASS system we adopt,\\
$A_{J} = 0.764E(B-V)$,\\
$A_{H} = 0.450E(B-V)$,\\
$A_{K_{s}} = 0.285E(B-V)$.\\
These values are from Laney \& Stobie (1993) as adjusted for $K_{s}$
by Gieren et al. (1998).
The table indicates the c-type variables. The fundamental periods
of these stars were obtained by multiplying the observed period
by 1.342.

\subsection{Results}
The revised Hipparcos parallax of RR Lyrae is $\pi = 3.46 \pm 0.64$.
Benedict et al. (2002a) found $\pi = 3.82 \pm 0.20$ from HST observations.
In the present paper we adopt a weighted mean of these values,
$\pi = 3.79 \pm 0.19$. 
This takes the quoted standard errors, each of which has their own
uncertainties, at their face value. Giving higher weight to the
globally-determined revised Hipparcos value would increase the
derived brightness of the star by $\leq 0.2$ mag.
We then obtain the following absolute magnitudes after adding a Lutz-Kelker
correction of --0.02 which was calculated on the same basis as that
adopted by Benedict et al.:\\
$M_{V} = + 0.54$, $M_{K_{s}} = -0.64$, \\
%$M_{K_{W}} = -0.80$,\\ 
each with standard error of $\pm 0.11$
In deriving the above figures we have adopted 
the data for RR Lyrae in Table~1. The reddening, $E(B-V) = 0.030$, 
given there 
agrees with the value derived directly from its
parallax distance and the Drimmel et al. (2003) formulation discussed 
below (0.031).

There are 142 stars, including RR Lyrae itself, in Table~\ref{142rr}.
%We have in fact omitted DH Peg in our final solutions since
%its status is uncertain. It is possibly a dwarf Cepheid
%(see Fernley et al. 1990).
Reduced parallax solutions (see, e.g. Feast 2002) were carried out
for this group of stars. 
%In each case the final reddenings adopted
%were obtained by iteration using the Drimmel et al. (2003) model. 
%A first estimate is made of the distance to the star assuming no
%interstellar extinction, the tabulated mean $K_{s}$ mag and\\
The reddenings were estimated for each star using the Drimmel et al.
(2003) three-dimensional Galactic extinction model, including
the rescaling factors that correct the dust column density to
account for small-scale structure seen in the DIRBE data but not
described explicitly by the model.  Two initial estimates were
made of the distance of a star using the tabulated mean
$K_{s}$ or $V$ magnitudes and preliminary PL($K_{s}$) or
$M_{V}-[Fe/H]$ relations, both of which correspond to an LMC modulus
of $\sim 18.5$. The results were iterated (see e.g. Whitelock et al.
(2008)). The values of $E(B-V)$ tabulated and used are the means of the
final results from $K_{s}$ and $V$. 

A reduced parallax solution of eq. 1 for the 142 stars and adopting 
$a = 0.214$, then leads to:\\
 $M_{V} = +0.54$,\\
 at the mean metallicity of the sample ($\overline{[Fe/H]} = -1.38$).
Similarly, reduced parallax solutions lead to,\\
$M_{K_{s}} = -0.63$\\ 
%and  $M_{K_{W}} = -0.81$,\\ 
at the mean $\log P$ of the sample 
($\overline{ \log P} = -0.252$), adopting a PL$(K_{s})$ slope
of --2.41 as in eq. 4. 
%and a PL($K_{W}$) slope of --2.51
%as in solution 1 (Table~1) of eq 5.
The standard error of  these derived absolute magnitudes
is $\pm 0.10$. 
(Note that no Lutz-Kelker correction is required in this case).
These results are essentially
identical to those for RR Lyrae itself and indeed the solution
is completely dominated by this one star. Omitting RR Lyrae
leads to solutions with very large standard errors. 
In the following we simply use the results based on RR Lyrae alone,
but using the full set of stars would obviously make no difference.

We then find,\\ 
$ b = 19.39 -Mod(LMC) = +0.84 \pm 0.11$\\
for eq. 1  with $a = 0.214$ as in eq. 2. This gives absolute
magnitudes brighter by $0.16 \pm 0.12$ than those given by
eq. 2 with an LMC modulus of $18.39 \pm 0.05$.
The standard error does not take into account the scatter
about the $M_{V}$ - [Fe/H] relation, which can be substantial
(see e.g. Gratton et al. fig. 19.).
This result is consistent with the prediction of
Catalan \& Cort\'{e}s (2008) that RR Lyrae is over luminous for its
metallicity by $0.06 \pm 0.01$mag compared with the average members
of this class. Note that if we adopted their preferred reddening for RR
Lyrae we would reduce the over luminosity implied by our result from
$0.16\pm 0.12$ to $0.12\pm 0.12$.
 
Main-sequence fitting procedures (Gratton et al. 2003) lead to
$b= +0.89 \pm 0.07$. However, other work (e.g. Salaris et  al. 2007)
has suggested a smaller distance modulus for 47 Tuc, 
a cluster on which the result of
Gratton et al. partly depends. Thus their value of $b$ may need 
increasing slightly.
The statistical parallaxes from Popowski \& Gould (1998) lead to
a value of $b = +1.10 \pm 0.12$, that is to absolute magnitudes
$0.10 \pm 0.12$ fainter than eq. 2. 

The parallax data on RR Lyrae leads to a constant term in eq. 3 of $-1.12$.
This is 0.07 mag brighter than the value given by Sollima et al. which was
based on the HST parallax of RR Lyrae alone and a slightly different $K_{s}$
magnitude. Following the discussion in Sollima et al. (2006), which takes
into account metallicities of the LMC variables, the parallax result leads
to a distance modulus of the LMC which is $0.22 \pm 0.14$ larger than that
deduced from the classical Cepheids ($18.39 \pm 0.05$).  A main uncertainty
in the Cepheid result was in the metallicity correction adopted, and the RR
Lyrae parallax result may indicate that this was overestimated. However, the
errors are such that within the uncertainties the classical Cepheids and RR
Lyrae variable scales are substantially in agreement.

%TABLE 2 ******
\begin{table}
\centering
\caption{Data for Type II Cepheids: Hipparcos Parallaxes}
\label{IICep_hip}
\begin{tabular}{lll}
\hline
  & VY Pyx & $\kappa$ Pav\\
\hline
$\log P $ &0.093 &0.959 \\
$[Fe/H]$ &--0.44  & 0.0 \\
$B$ &7.85  & 4.98 \\
$V$ &7.30  & 4.35\\
$I$ &  & 3.67\\
$J$ &6.00  & 3.17 \\
$K_{s}$ &5.65  &2.78\\
$E(B-V)$ &0.049 & 0.017\\
$\pi$ &5.00 & 6.51\\
$\sigma_{\pi}$ &0.44 & 0.77\\
$ Mod $ &6.59 & 5.93\\
$\sigma_{Mod} $ & 0.19 & 0.26\\
$ LK $ &--0.06 & --0.12\\
$M_{B}$ &+1.09 & --1.14\\
$M_{V}$ & +0.54 & --1.86\\
$M_{I}$ & & --2.41\\
%$M_{V_{W}}$ & & --3.06\\
$M_{K_{s}}$ &--0.92 & --3.27\\
%$M_{K_{W}}$ &-1.12 & -3.50\\
\hline
\end{tabular}
\end{table}

\section{The Type II Cepheids}
\subsection{Trigonometrical parallaxes}
 The relevant data for the two CephIIs 
on our programme are collected in Table~2.
The metallicity of VY Pyx is from Maas et al. (2007).
The value quoted for $\kappa$ Pav is from
Luck \& Bond (1989). Both stars are
comparatively metal-rich. The $BV$ photometry of VY Pyx is from
Sanwal \& Sarma (1991), whilst $J$ and $K_{s}$ are single 2MASS values. 
In view of the low
visual amplitude of VY Pyx ($\Delta V = 0.27$), these should be close
to mean values.
The magnitudes, light curve and period agree satisfactorily with the
Hipparcos photometry (ESA 1997).
For $\kappa$ Pav the intensity mean $B$, $V$ and $I$ were derived from
from the literature cited in Table~3, 
with $I$ in the Cousins system. 
$J,K_{s}$ for this star are from the intensity means given in section 4.2.2
transformed to the 2MASS system using the relations derived by
Carpenter (2001 as updated on the 2MASS Web page). The reddenings for 
both stars were
estimated on the Drimmel et al. (2003) model described in section 3.2, 
with distances
adopted from the revised Hipparcos parallaxes ($\pi \pm \sigma_{\pi}$)
which are also listed. The distance moduli ($Mod$) and their uncertainties
come directly from the parallaxes. The Lutz-Kelker ($LK$) corrections
needed in deriving the absolute magnitudes are calculated on the same
system as used for RR Lyrae (section 3). In discussing the various
absolute magnitudes listed we shall use for their standard errors the
values derived for the distance moduli. It should be borne in mind
that these may be slightly underestimated due to any uncertainty in
photometry, reddening and Lutz-Kelker correction.

%TABLE 3 ******
\begin{table*}
\centering
\caption{Pulsation Parallax solutions for Classical Cepheids and $\kappa$ Pav} 
\begin{tabular}{llccccccccc}
\hline
Star & Period & $<K_{o}>$ & $<J_{o}>$ & $<V_{o}>$ & $R_{1}$ & $R_{2}$ &
$M_{K}$ & $\pi_{1}$ & $\pi_{2}$ & $p$ \\
\hline
$\delta$ Cep & 5.3662475 & 2.295 & 2.678 & 3.667 & 
$41.3 \pm 1.0$ &  $42.5 \pm 1.0$ & $-4.86$ &
$3.71 \pm .12$ & $3.72 \pm .09$ & $1.27 \pm .05$ \\  
X Sgr & 7.012675 & 2.453 & 2.833 & 3.819 & $49.3 \pm 1.6$ & 
$47.3 \pm 1.4$ & $-5.16$ & $3.17 \pm .14$ & $3.01 \pm .09$ & 
$1.20 \pm .06$ \\
$\beta$ Dor & 9.842578 & 1.947 & 2.405 & 3.616 & $62.1 \pm 1.7$ &
$63.0 \pm 1.0$ & $-5.64$ & $3.26 \pm .14$ & $3.04 \pm .07$ &  $1.18 \pm .06$ \\
$\zeta$ Gem & 10.14992 & 2.128 & 2.605 & 3.884 & $62.7 \pm 1.7$ &
$65.4 \pm 1.6$ & $-5.67$ & $2.74 \pm .12$ & $2.76 \pm .07$ & 
$1.28 \pm .06$ \\
$l$ Car & 35.54327 & 1.046 & 1.639 & 3.225 & $162.3 \pm 4.0$ &
$165.7 \pm 3.0$ & $-7.59$ & $2.03 \pm .16$ & $1.87 \pm .04$ & 
$1.17 \pm .10$ \\
$\kappa$ Pav & 9.0880 & 2.795 & 3.201 & 4.291 & $26.5 \pm 0.8$ &
$26.3 \pm 0.6$ &  $-3.81$ & $6.51 \pm .77$ & $4.78  \pm .13$ & 
$0.93 \pm .11$ \\
\hline
\end{tabular}
The columns contain: (1) star name, (2) period in days, (3,4,5) intensity
mean magnitudes corrected for reddening ($<K_{o}>$, $<J_{o}>$ in the SAAO
system), (6,7) radii in solar units derived from $K$, $J-K$ ($R_{1}$) and
$K$, $V-K$ ($R_{2}$) with $p = 1.27$ (8) the trigonometrical parallax and
its s.e. (9) pulsation parallax and its (internal) s.e. (11) $p$. The errors
of the mean radius and the trig. parallax have been added in quadrature
for $\sigma_{p}$. \\
References: $\delta$ Cep, 1,2,3,A,B,C; X Sgr, 1,4,5,6, D-N; $\beta$ Dor, 7-10,
O,P; $\zeta$ Gem, 1,3,13,A,C,Q; $l$ Car, 7,9,10,11,M.R; $\kappa$ Pav,
7,9,13,15,16,P.\\ Optical Photometry references: (1) Moffett \& Barnes 1984,
(2) Barnes et al. 1997, (3) Kiss 1998, (4) Shobbrook 1992, (5) Arellano Ferro
et al. 1998, (6) Berdnikov \& Turner 2001, (7) Dean et al. 1977, (8) Pel
1976, (9) Dean 1981, (10) Shobbrook 1992, (11) Bersier 2002, (12) Szabados
1981, (13) Dean 1977, (14) Berdnikov 1997, (15) ESA 1997, (16) Cousins \&
Lagerweij 1971.

Radial Velocity references: (A) Bersier et al. 1994, (B) Butler 1993, (C)
Kiss 1998, (D) Moore 1909, (E) Duncan 1932, (F) Stibbs 1955, (G) Feast 1967,
(H) Lloyd Evans 1968, (I) Lloyd Evans 1980, (J) Barnes et al. 1987, (K)
Wilson et al. 1989, (L) Sasselov \& Lester 1990, (M) Bersier 2002, (N)
Mathias et al. 2006, (O) Taylor \& Booth 1998, (P) Wallerstein et al. 1992,
(Q) Gorynya et al. 1998, (R) Taylor et al. 1997.
\end{table*}

There are other stars classified as CephIIs in the Hipparcos catalogue
in addition to $\kappa$ Pav and VY Pyx, but their $\sigma_{\pi} / \pi$
values are relatively high  and in some cases it is uncertain whether
they belong to the CephII class. We have therefore not attempted
to use these stars.
\subsection{Pulsation parallaxes}
\subsubsection{The Projection factor, $p$}
 The Baade-Wesselink method for radius determination has seen only limited
use for CephIIs, even at optical wavelengths, and table~2 in Balog et al.
(1997) suggests that such results as have been reported are somewhat
inconsistent with each other.

For classical Cepheids, the reasons for using IR photometry in determining
pulsation parallaxes or Baade-Wesselink radii have been given by Laney \&
Stobie (1995 henceforth LS95), and by Gieren, Fouqu\'{e} \& Gomez (1997),
among others. This technique has not been used previously in determining
radii, luminosities, etc. for CephIIs, except for a few preliminary results
given by Laney (1995). Whilst modern pulsation parallaxes are often of high
internal consistency, it has been difficult to estimate possible systematic
uncertainties.  Significant progress in dealing with such systematic
uncertainties has become possible since the advent of reasonably accurate
parallaxes for nearby classical Cepheids (Benedict et al. 2002b, 2007, van
Leeuwen et al. 2007), as these allow a particular pulsation parallax method
to be calibrated empirically.

Several recent papers (Merand et al. 2005, Groenewegen 2007, Nardetto et al.
2007, Fouqu\'{e} et al. 2007) have tackled the determination of the
projection factor ($p$-factor), which has long been one of the principal
sources of uncertainty in pulsation parallaxes. Other papers have discussed
angular diameter measurements and the surface-brightness colour relation,
but these are not as directly relevant to the method used here, as the radii
derived in this paper have been calculated using the technique described in
Balona (1977), where the surface-brightness coefficient is a free parameter.
Conversion from radii to luminosities uses a methodology described below,
and is in effect included in the calibration of the $p$-factor.

As in  LS95, solutions have been derived with a modified version of Luis
Balona's software which allows for a non-negligible amplitude, and where
photometric magnitudes and colours, as well as radial velocities, are
assigned individual errors. All radii used were derived using $K$ as the
magnitude and $V-K$ or $J-K$ as the colour, as this approach was shown to be
free of serious phase-dependent systematic error by LS95.  
These authors also show that inclusion or exclusion of the rising branch
has a negligible systematic effect on the derived radii, although
excluding the rising branch increases the uncertainty in the results.
Here $J,K$ are on
the SAAO system (see below, Appendix A).
Adopted radii are the means of the ($K,V-K$)
and ($K,J-K$) values. The adopted formal error in the radius is derived by
taking the square root of the mean of the squares of the individual errors
in the ($K,V-K$) and ($K,J-K$) radii.

The first necessary step is to derive an appropriate value of the $p$-factor
{\it for the specific method used here}. Our radius-determination
methodology is different from those used 
by Merand et al. (2005), Groenewegen (2007) and Nardetto et al. (2007),
and the radial velocities (selected from the literature) are not
based on a single selected line, as described by Nardetto et al. (2007).

As a first approximation, $p=1.27$ (Merand et al. 2005, Groenewegen 2007) was
adopted, and radii were calculated for five of the classical Cepheids
in table~2 of van Leeuwen et al. 2007).
Polaris has a limited, variable amplitude and we are unaware of
suitable data for an accurate radius solution. For FF Aql the possible
influence of a binary companion and the low quality of the $JHK$ data were
enough to drop it from the list. The other stars in the van Leeuwen et al.
list have higher $\sigma_{\pi}/\pi$ than our five stars.

\begin{figure}
\includegraphics[width=8.5cm]{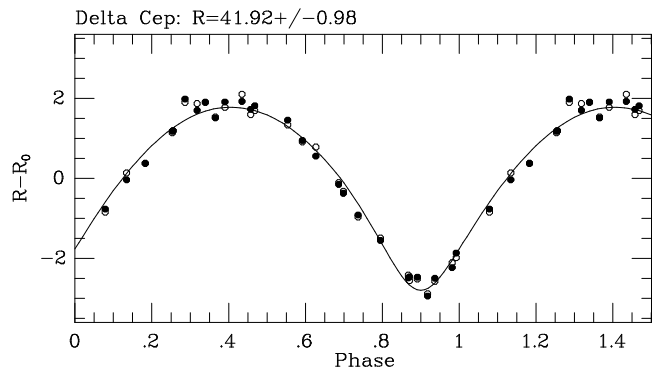}
\caption{Radius displacements for $\delta$ Cep calculated from the
$K$, $J-K$ (open circles) and $K$, $V-K$ (filled circles) radius solutions
and photometry, vs. the integrated radial velocity curve (solid line).
A projection factor of $p$ = 1.27 was used.}\label{L1}
\end{figure}

\begin{figure}
\includegraphics[width=8.5cm]{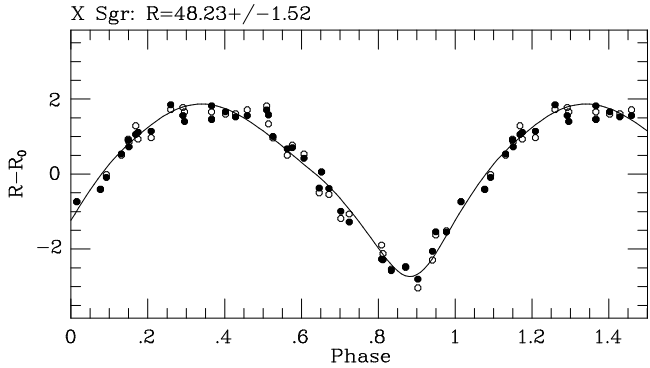}
\caption{As Fig.~\ref{L1}, but for X Sgr.}\label{L2}
\end{figure}

\begin{figure}
\includegraphics[width=8.5cm]{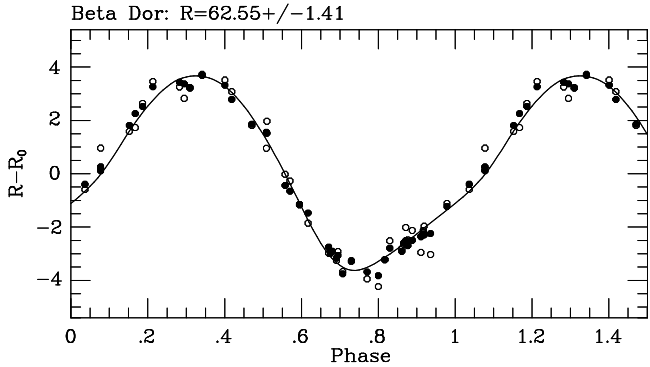}
\caption{As Fig.~\ref{L1}, but for $\beta$ Dor.}\label{L3}
\end{figure}

\begin{figure}
\includegraphics[width=8.5cm]{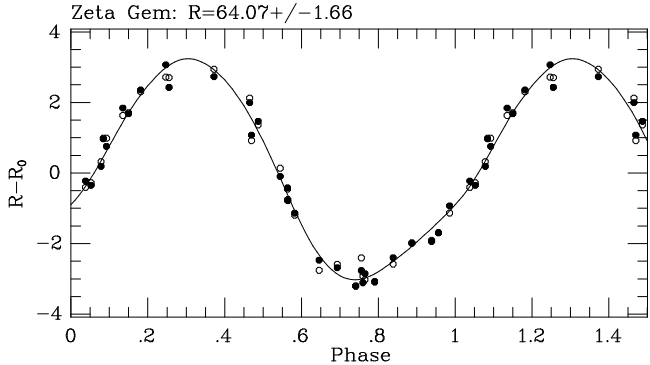}
\caption{As Fig.~\ref{L1}, but for $\zeta$ Gem.}\label{L4}
\end{figure}

\begin{figure}
\includegraphics[width=8.5cm]{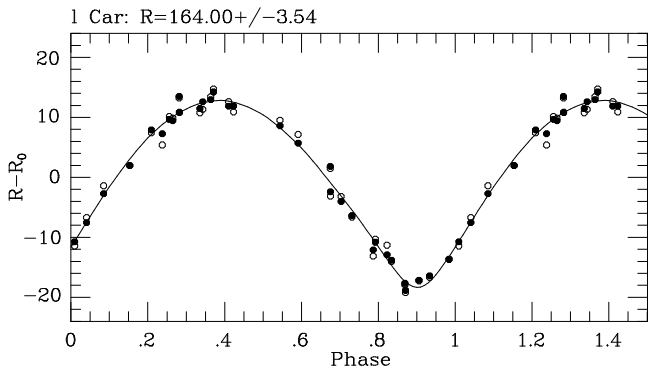}
\caption{As Fig.~\ref{L1}, but for {\it l} Car.}\label{L5}
\end{figure}

\begin{figure}
\includegraphics[width=8.5cm]{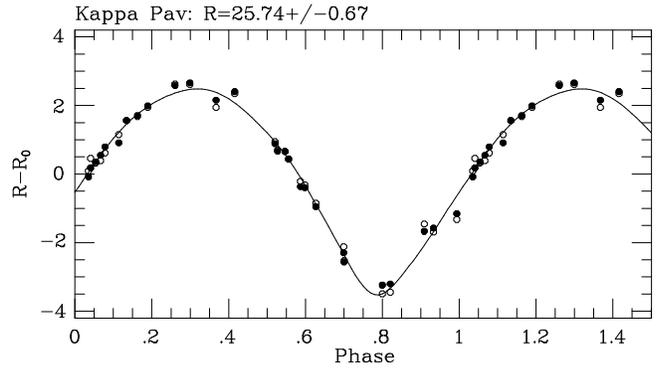}
\caption{As Fig.~\ref{L1}, but for $\kappa$ Pav.}\label{L6}
\end{figure}

For the remaining five stars, the ($K$, $V-K$) and ($K$, $J-K$) radii were
calculated with $p=1.27$, then converted into luminosities. This was done
using the tables given in Hindsley \& Bell (1990) to establish the $K$-band
absolute magnitudes for a star of one solar radius and the appropriate
dereddened $V-K$ and $J-K$ colours, then taking the mean. 
As discussed in LS95, the $K$ surface brightness as a function of $J-K$
or $V-K$ is very insensitive to surface gravity or microturbulence, which means
that neither the radius solution nor the derived luminosity is
significantly affected by assumptions about mean or time-varying values
for these quantities in the stellar atmosphere.
A similar
procedure was followed for $\kappa$ Pav, the only CephII which has
good $JHK$ 
and radial velocity data and a usable parallax measurement -- though 
this is
of lower quality than 
for the five classical Cepheids. Dereddening was done
using the reddening coefficients derived by Laney \& Stobie (1993), and $BVIc$
reddenings for each star as calibrated recently by Laney \& Caldwell (2007),
using metal abundances from the tables in that paper, or for $\kappa$
Pav the value from Luck \& Bond (1989). The resulting small uncertainty
in the colours has only a small effect on the $K$ surface brightness, as it is
only a weak function of either $V-K$ or $J-K$. Figs.~\ref{L1}-\ref{L6} show
the match of radius displacements calculated from the radius solution and
$VJK$ photometry to the integrated radial velocity curve. As would be expected
from LS95, there are no serious phase anomalies or discrepancies. Any
serious problems with shock waves, etc. that distorted the solutions should
appear in these diagrams, but there is no real sign of such an effect --
even for X Sgr (Sasselov \& Lester 1990, Mathias et al. 2006) or $\kappa$
Pav. For any other value of the projection factor, the curves would appear
identical to those shown except that the vertical scale would be
slightly different.

\begin{figure}
\includegraphics[width=8.5cm]{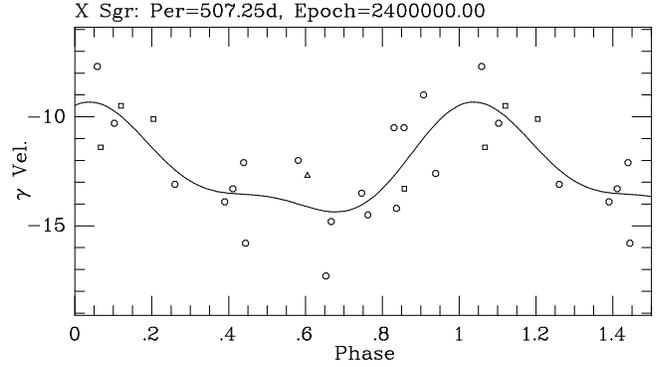}
\caption{Gamma velocities for X Sgr, phased according to the ephemeris
and period of Szabados (1990). The squares represent data from Bersier
(2002) and Sasselov \& Lester (1990). The triangle is the value from
Mathias et al. (2006).}\label{L7}
\end{figure}

\begin{figure}
\includegraphics[width=8.5cm]{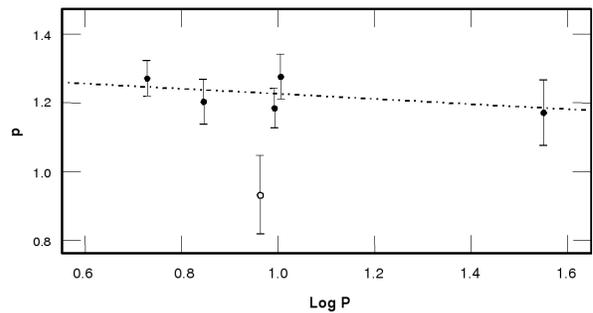}
\caption{ The projection factor, $p$, plotted against $\log P$ for the
classical Cepheids, $\delta$ Cep, X Sgr, $\beta$ Dor, $\zeta$ Gem,
and {\it l} Car (filled circles) and the CephII $\kappa$ Pav
(open circle). The line shows the trend of $p$ with period suggested by
Nardetto et al. (2007), but adjusted to the zero-point given by
the five classical Cepheids.}\label{L8}
\end{figure}

\begin{figure}
\includegraphics[width=8.5cm]{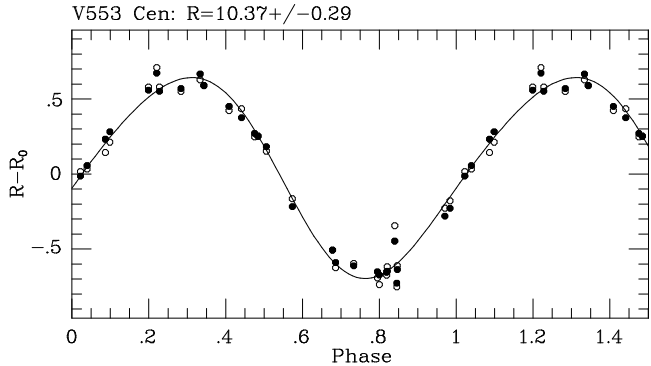}
\caption{As Fig.~\ref{L1}, but for V553 Cen and adopting $p$ = 1.23.}\label{L9}
\end{figure}

\begin{figure}
\includegraphics[width=8.5cm]{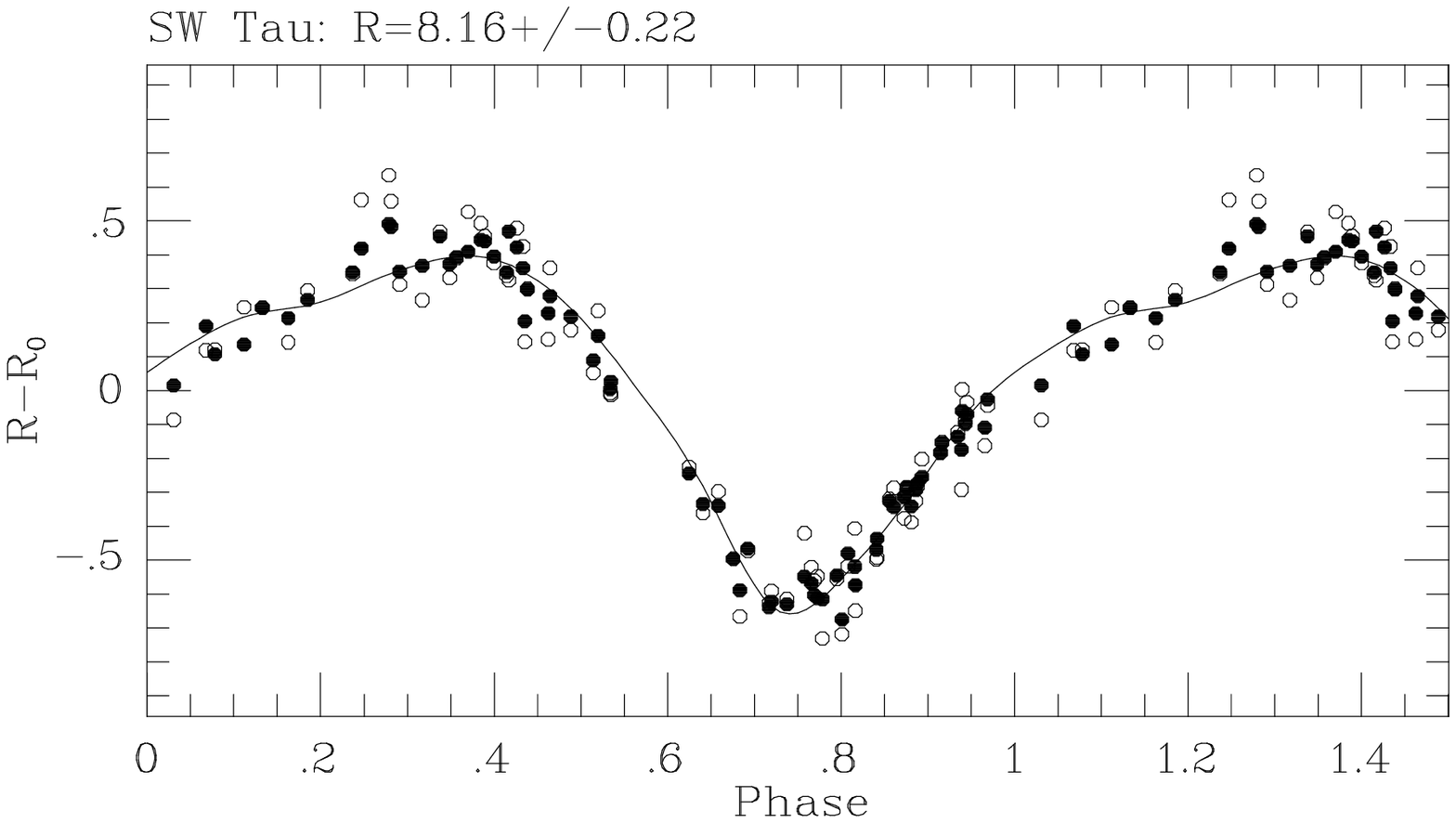}
\caption{As Fig~\ref{L1}, but for SW Tau and adopting $p$ = 1.23.}\label{L10} 
\end{figure}

In all cases, it was necessary to establish the phase and period behaviour
of the star, so that there were no systematic shifts between the phases or
zero-points of the optical photometry, infrared photometry and radial
velocities. For X Sgr, it was also necessary to redetermine the orbital
velocity curve, in view of the doubts expressed by Mathias et al. (2006).
All velocities in the literature for this star, including the most recent,
appear to be consistent with the orbital period determined by Szabados
(1990), and it proved possible to separate the orbital and pulsational
velocities effectively (Fig. 7), though better data are desirable.
The $JHK$ data used are listed in Appendix A.

%\begin{figure*}
%\includegraphics[width=8.5cm]{tableL2.ps}
%\caption{Table 2}\label{table2} 
%\end{figure*}

Radii, luminosities 
and pulsation parallaxes for the five classical Cepheids and $\kappa$ Pav,
derived as above for $p=1.27$, are given in Table~3, together with the
sources for the optical photometry and radial velocities. 
Also in this table are the trigonometrical parallaxes from
van Leeuwen et al. (2007)
and the present paper. 
Requiring that the $p$-factor for each star be adjusted to produce agreement
between the pulsation  and trigonometric parallaxes leads to the empirical
$p$-factors for each star listed in Table 3 together with  the
associated errors due to the uncertainties in both the radius and the
trigonometric parallax.
These lead to the empirical $p$-factor for
each star listed in the table
%These were then
%used together with the parallaxes and Lutz-Kelker corrections from van
%Leeuwen et al. (2007) to derive the empirical p-factor for each star given
%in Table 3 
together with the associated errors due to uncertainty in the radius and in
the parallax. These values of $p$ are plotted against $\log P$ in Fig. 8.
For all 5 classical Cepheids, the derived $p$-factor falls within a narrow
range, and the mean is $1.22\pm 0.02$, weighting the stars equally. An
average, weighted according to the inverse square of the error, gives
$1.23\pm 0.03$
where the weight of $l$ Car has been set to one and its error has
been divided by the square root of the sum of the weights for all
five stars . 
A trend with period may be present, as claimed in Nardetto
et al. (2007), though our sample is too small to derive a useful,
statistically significant value of a term in $\log P$. If we assume that
there is a $\log P$ term of --0.075 (given by Nardetto et al. as appropriate
for velocities based on a mix of lines of varying depth), the 
weighted intercept at $\log P = 1.0$ is $1.23 \pm 0.03$.

%zero-point would be
%$1.312\pm0.017$, and the weighted zero-point 1.311.
%TABLE 4 ******
\begin{table*}
\centering
\caption{Pulsation Parallax Results for Type II Cepheids}
\begin{tabular}{llccccccc}
\hline
Star & Per & $<K_{o}>$ & $<H_{o}>$ & $<J_{o}>$ & $<V_{o}>$ &
$R_{1}$ & $R_{2}$ & $D$ \\
\hline
SW Tau & 1.583565 & 7.887 & 7.931 & 8.147  & 8.800 &
$8.02 \pm 0.27$ & $8.03 \pm 0.15$ & $732\pm 20 \pm 16$ \\
V553 Cen & 2.060464 & 6.878 & 6.963 & 7.290 & 8.455 & $10.53 \pm 0.33$
& $10.20 \pm 0.25$ & $541 \pm 15 \pm 12$\\
$\kappa$ Pav & $9.0902$ & 2.795 & 2.863 & 3.201 & 4.291 &
$26.48 \pm 0.78$ & $26.32 \pm 0.62$ & $204 \pm 5 \pm 4$ \\
\hline
\end{tabular}
 
The columns are: (1) star name, (2) period in days (for $\kappa$ Pav
this is the mean of the three periods used for the optical photometry),
(3,4,5,6) intensity mean magnitudes with the infrared values on the
SAAO system, (7,8) radii in solar units from, $K,J-K$ ($R_{1}$)
and $K,V-K$ ($R_{2}$), 
(9) distance in pc based on a mean of $R_{1}$ and $R_{2}$ and with
$p=1.23$ (the first standard error reflects the uncertainty in
the derived radius, the second the uncertainty in $p$).
\end{table*}

The derived $p$-factor for $\kappa$ Pav, on the other hand, is strikingly
discrepant, so low as to be physically unrealistic, especially given that
the colours and surface gravity are in reasonable accord with those given
for classical Cepheids by Laney \& Stobie (1994) and Fernie (1995)
respectively, while the metallicity is solar (Luck \& Bond 1989) and the
radius displacement diagram (Fig.~\ref{L6}) resembles those of the 5
classical Cepheids. However, the parallax for this star is more uncertain
than for the five classical Cepheids, and the derived $p$-factor is in fact
only about 2 $\sigma$ from the weighted mean of the
5 classical Cepheids.
%$1.228\pm 0.027$ for all six
%stars (a mean hardly affected by $\kappa$ Pav because of its low
%statistical weight). 
A $p$-factor of 1.23 was adopted for all three CephIIs
considered here\footnote{ See also the discussion in section 5.1.}.
Details of the radius and luminosity determinations follow. Magnitudes,
radii, absolute magnitudes 
and other relevant data are given in Tables~4 and 5.

\subsubsection{$\kappa$ Pav}

The best-fitting period for the IR data in Table A1 (JD 2445928-2447769)
was 9.0814 d, and the scatter around a low-order (2 to 5) Fourier fit to the
resulting magnitudes and colours was about 0.009-0.011 mag. This is rather
higher than normal for such a bright star, and suggests a modest amount of
phase jitter may have been present.

Contemporaneous radial velocity data were available in the literature
(Wallerstein et al. 1992), covering almost exactly the same range of Julian
dates. A modest number of velocities with slightly later JD were shifted
into phase agreement at the adopted period. The light curve of $\kappa$ Pav
is known for sudden changes (Wallerstein et al. 1992), so a need for phase
adjustments is not surprising. 

The sources of the visual photometry are given in Table~3. All datasets have
been phased at their appropriate periods, then shifted into phase and
zero-point agreement with Dean et al. (1977) and Dean (1981). This composite
dataset was used to derive a 6th order Fourier fit to the $V$ light curve,
with maximum light in $V$ set to phase 0. None of the optical photometry
data sets was contemporary with the infrared data. Derived periods and
epochs were:\\
2440140.119   + 9.0947E   (Cousins \& Lagerweij)\\
2441959.49903 + 9.08352E  (Dean et al., Dean)\\
2448164.8647  + 9.092405E (Shobbrook, Hipparcos, Berdnikov, Berdnikov \&
                           Turner)

A $V$ magnitude was then calculated for each infrared observation, using an
epoch for the IR data which ensured that a Fourier fit to the $V-K$, 
and $J-K$
data gave phases for minimum light in agreement with those for $B-V$ and
$V-I$, a technique for phase alignment validated by LS95. The resulting 
($K,J-K$) and ($K,V-K$) radii agree within less than one percent, and there
are no significant phase-dependent anomalies (Fig.~\ref{L6}).

$E(B-V)=0.017\pm 0.022$ was derived from the $B-V$ and $V-I$ magnitude means (and
the solar metallicity given by Luck \& Bond (1989)), using the Cousins
reddening method as re-calibrated by 
Laney \& Caldwell (2007). While this method has not been
specifically calibrated for CephIIs, $\kappa$ Pav falls into much the
same range in temperature, surface gravity and metallicity as classical
Cepheids. This reddening is virtually the same as that derived by the
Drimmel method (0.019). 
The reddening value is in any event not critical -- it affects the
luminosity and distance determinations {\it only} through the weak
dependence of $K$ surface brightness on the dereddened $V-K$ and $J-K$ colour
indices.

Dereddened $V-K$ and $J-K$ colours were used to calculate the surface brightness
at $K$ as described above, using $\log g$ of 1.2 (Luck \& Bond 1989), and
converted to absolute magnitudes at $V, J$ and $K$ using the mean radius and the
dereddened empirical colours. 2MASS $J$, $H$ and $K_{s}$, 
absolute magnitudes were
calculated using the transformations on the 2MASS website, as they also were
for V553 Cen and SW Tau, below.

%\subsection{Type II Cepheids in the field and in the clusters NGC 6441 
%and 6388}
%TABLE 5 ******
\begin{table}
\centering
\caption{ Data for Type II Cepheids: Pulsation Parallaxes}
\begin{tabular}{llll}
\hline
&  $\kappa$ Pav & V553 Cen &  SW Tau\\
\hline
$\log P$ & 0.959 & 0.314 & 0.200 \\
$[Fe/H]$ & 0.0  & +0.24 & +0.22 \\
$B$    & 4.98  & 9.15 & 10.32 \\
$V$    & 4.35  & 8.46 & 9.66 \\
$I$   & 3.67 & 7.76 & 8.94\\
$K_{s}$ & 2.78 & 6.86 & 7.95\\
$K_{W}$ & 2.55 & 6.63 & 7.73 \\
$E(B-V)$ & 0.017 & 0.00 & 0.282 \\
$Mod$ & 6.55 & 8.67 & 9.32\\
$\sigma_{Mod}$ & 0.07 & 0.08 & 0.08\\
$M_{B}$ & --1.64 & +0.48 & --0.15\\
$M_{V}$ & --2.25 & --0.21 & --0.53\\
$M_{I}$ & --2.91 & --0.90 & --0.88\\
$M_{K_{s}}$ & --3.77 & --1.80 & --1.46 \\
%$M_{K_{W}}$ & -4.00 & -2.03 & -1.59\\
\hline
\end{tabular}
\end{table} 

\subsubsection{V553 Cen}

The period behaviour is simpler than for $\kappa$ Pav, and seems
adequately described by:

2448437.1154 + 2.060464E (2444423-2450364)\\
2443108.6572 + 2.060608E (2440700-2443686)\\

These phases were adopted for the IR photometry (Table A1), for 
optical photometry by Wisse \& Wisse (1970), Lloyd Evans et al. (1972), 
Dean et al. (1977), Dean (1981), Eggen (1985), Diethelm (1986), Gray \&
Olsen (1991), ESA (1997), Berdnikov \& Turner (1995) and Berdnikov
(1997), and for radial velocities by Wallerstein \& Gonzalez (1996) and
Lloyd Evans et al. (1972). All optical photometry was adjusted in zero point
to match Dean et al. (1977) and Dean (1981), and the radial velocities to
match Wallerstein and Gonzalez.

The mean $E(B-V)$ for solar metallicity and a microturbulence of 2.5 $\rm
km\,s^{-1}$ (Wallerstein and Gonzalez 1996) is $0.00\pm 0.02$ from 54
observations with $B-V$ and $V-I$.
These authors also derive $\log g \sim 1.8$. The
Drimmel procedure gives $E(B-V) = 0.08$.

The derived $(K,J-K)$ and $(K,J-K)$ radii agree within the errors, and the lack
of significant phase-dependent anomalies can be seen in Fig.~\ref{L9}.

\subsubsection{SW Tau}

The period seems essentially constant at 1.583565d over the relevant
interval, with an epoch of 2445013.2696 for maximum light in $V$. Optical
photometry has been taken from Barnes et al. (1997), Moffett \& Barnes
(1984), and Stobie \& Balona (1979), and 
the zero-point shifted to match Stobie
\& Balona. For $B-V$ and $V-I$ magnitude means of 0.653 and 0.796 on the
Cousins system, with $[Fe/H]=+0.2$, microturbulence of 3.0 $\rm km\,s^{-1}$
(Maas et al. 2007), $E(B-V)$ is $0.282\pm 0.031$. Log g from Maas et al. is
about 2.0. The Drimmel procedure gives $E(B-V) = 0.26$.

IR data for SW Tau on the CIT system was taken from Barnes et al. (1997) and
transformed to the Carter system by the formulae given in Laney \& Stobie
(1993). This was then combined with the SAAO $JHK$ observations, and matched
to the SAAO zero point. As would be expected, the resulting shifts were
small. 

Radial velocities used are those from Gorynya et al. (1998) and from Bersier
et al. (1994).

The derived ($K,J-K$) and ($K,J-K$) radii agree within the errors, and the lack
of significant phase-dependent anomalies can be seen in Fig.~\ref{L10}.

\section{Discussion}
\subsection{$\kappa$ Pav}
\begin{figure}
\includegraphics[width=8.5cm]{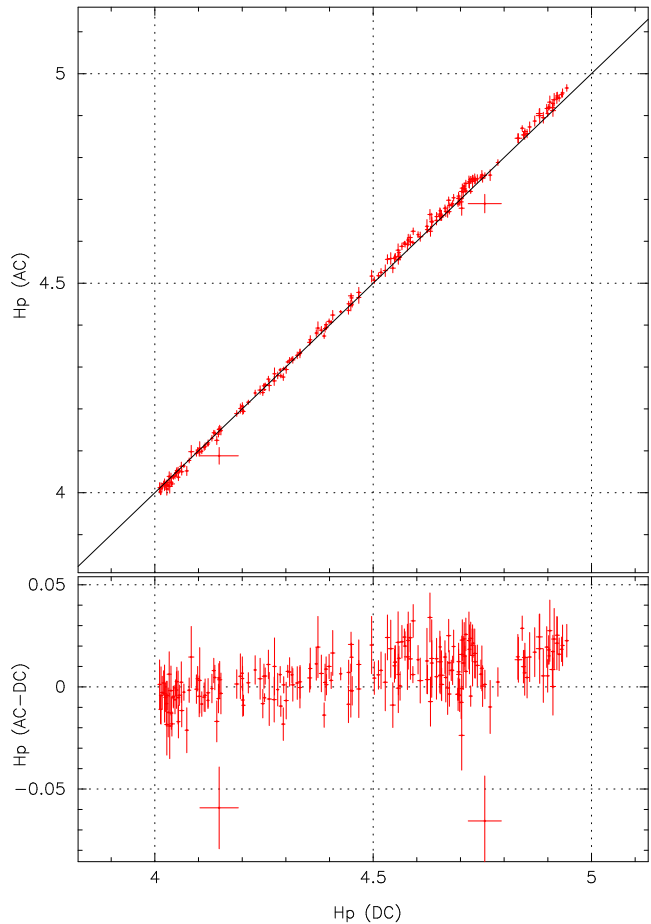}
\caption{The relation between the Hipparcos AC and DC magnitudes for
$\kappa$ Pav. The increasing discrepancy between the AC and DC magnitudes
towards fainter magnitudes is an indication for the presence of a close
companion that becomes more visible as $\kappa$ Pav becomes fainter.}
\label{vL}
\end{figure}

  The trigonometrical and pulsational parallaxes of $\kappa$ Pav
are $6.51 \pm 0.77$ and $4.90 \pm 0.17$, a difference of
$1.61 \pm 0.79$. This 2$\sigma$ difference is sufficiently large
to raise some concerns. The Hipparcos result is from a type 3 solution. In
such a solution account is taken of possible variability induced motion.
Further investigation shows evidence (Fig. \ref{vL}) for a magnitude
dependence difference between the DC and AC Hipparcos magnitudes. These
magnitude systems and the interpretation of differences between them are
described in the Hipparcos catalogue (ESA 1997). The results for $\kappa$ Pav
suggest the presence of a close companion consistent with the need for a
type 3 solution. Given the method
of reduction employed, the revised Hipparcos parallax should be reliable
within the quoted uncertainty. 

The possibility that $\kappa$ Pav
was a spectroscopic binary was suggested by Wallerstein et al. (1992)  
from a comparison of their work with much earlier observations. There is,
however, no evidence of short period variations in $\gamma$ velocity in 
their data
which extended over a considerable time span (JD 2445860-2448283)
or the additional data we have used.
%and on which the present analysis depends.
The five-colour photometry of Janot-Pacheco (1976) shows no evidence
of a bright companion. The present work provides internal checks
on the possibility of a bright companion. A bright red companion
would produce abnormally low surface brightness coefficients in the
($K,J-K$) and, especially, the ($K,V-K$) solutions. A companion
of similar colour to the variable would affect the two solutions more
equally. In fact, these two surface brightness coefficients are
slightly higher for $\kappa$ Pav than the other two CephIIs in the
programme, though not significantly so.  A blue companion would
tend to make the ($K,V-K$) radius smaller than the ($K,J-K$) one.
The ($V,B-V$) radius would be smaller still and have an unusually
large surface brightness coefficient as seen in the classical
Cepheid binary KN Cen (LS95). In $\kappa$ Pav
there is no significant difference between the ($K,V-K$) and
($K,J-K$) radii. The ($V,B-V$) radius is smaller by 13 percent.
This is a marginal effect and indicates that any blue companion
has a relative brightness considerably fainter than in the
case of KN Cen. 

Thus, in summary, no serious anomalies were found in the
pulsation parallax analysis besides the problem of phase shifts. 
However, some caution is necessary in discussing this star. In the
following, we discuss the results separately for the two estimates of the
parallax.

\subsection{Infrared period-luminosity relations}
 Table~\ref{irpl}  lists the differences of the parallax based absolute
magnitudes from the PL($K_{s}$) relation (eq. 4).  We adopt $c= -1.0$
corresponding to an LMC modulus of 18.39. Besides the CephII stars,
Table~\ref{irpl} lists, in addition, the results for RR Lyrae. As already
noted, Matsunaga et al. (2006) suggested that the RR Lyrae variables lay on
the same PL($K_{s}$) relation as the CephIIs and this suggestion was
strengthened by the work of Sollima et al. (2006).  Two standard errors are
given, $\sigma_{1}$ is the value derived from the parallax solution and
$\sigma_{2}$ combines this
%The standard errors
%of the differences in Table~\ref{irpl} (7)were obtained by combining the
%standard errors of the moduli 
in quadrature with the scatter in the 
PL($K_{s}$)
relation as given by Matsunaga et al. (2006) (0.14). This latter value is an
upper limit to the intrinsic scatter of the Matsunaga et al. relation since
it includes uncertainties in the moduli of the globular clusters they used
etc.  The first part of Table~6 shows the results from the trigonometrical
parallaxes and the second part the results from the pulsation parallaxes.

%TABLE 6 *******
\begin{table}
\centering
\caption{Differences from Infrared PL relations}\label{irpl}
\begin{tabular}{rrrr}
\hline
Star & $\Delta M_{K}$ & $\sigma_{1}$ & $\sigma_{2} $\\
\hline
& & (a)&\\
\hline
RR Lyrae & --0.24 & 0.11 & 0.17\\
VY Pyx   & +0.30 & 0.19 & 0.24\\
$\kappa$ Pav & +0.04 & 0.26 & 0.30\\
\hline
& & (b) &\\
\hline
$\kappa$ Pav & --0.46 & 0.07 & 0.16\\
V533 Cen & --0.05 & 0.08& 0.16\\
SW Tau & +0.02 & 0.08 & 0.16\\
\hline
\end{tabular}

(a) Results using trigonometrical parallaxes.\\
(b) Results using pulsational parallaxes.\\
\end{table}

     Given the uncertainties in the trigonometrical parallaxes, the results
in the first part of Table~\ref{irpl} show satisfactory agreement with the
predictions of the infrared PL relation. The two short period stars with
pulsation parallaxes (SW Tau, P = 1.58; V553 Cen, P =2.06) agree closely
with predictions. This agreement is sufficiently good to hint that the
intrinsic scatter in the relations is less than the adopted 0.14, in
agreement with the discussion above.  Indeed if the possible period
dependence of the projection factor $p$ discussed in section 4.2.1 applies,
these two stars lie even more closely on a line with the Matsunaga et al.
slope. They would then be 0.09 mag (V553 Cen) and 0.08 mag (SW Tau) brighter
than that predicted using a zero-point based on an LMC modulus of 18.39.
Both SW Tau and V553 Cen are carbon-rich stars of near solar metallicity. SW
Tau has [Fe/H] = +0.22 (Maas et al. 2007) and V553 Cen has [Fe/H] = +0.04
(Wallerstein \& Gonzales 1996). The light-curve classification scheme
proposed by Diethelm (1990 and other papers referenced there) indicates
that, as one would expect, these two stars are disc objects. On the other
hand the short-period globular-cluster stars ($P < 5$ days) in Matsunaga et
al. (2006) are all of low metallicity ([Fe/H] in the range --1.15 to
--1.94). Thus within the uncertainties, the PL($K_{s}$) relation for CephIIs
is insensitive to population differences (metallicity, mass) 
at least at the short period end.

 The pulsation parallax of $\kappa$ Pav leads to an infrared
absolute magnitude that differs significantly from the PL relations
derived from the globular clusters and with an LMC modulus of 18.39.
Since the formal uncertainty of the pulsation-based absolute magnitude is
0.07 mag the deviation is $6.5\sigma$ and even taking into account
the upper limit on the intrinsic scatter in the PL($K_{s}$) relation
there is nearly a three sigma deviation.  Evidently if this result
is accepted then some CephIIs in the field can deviate significantly from
the PL($K_{s}$)
based on globular cluster variables. Since the metallicity of 
$\kappa$ Pav is near solar and the results of Matsunaga et al. depend
on metal-poor objects,
a (large) metallicity effect might be the cause. As
there is little metallicity dependence among the metal-poor objects
(see section  2.2) this would imply a very non-linear dependence
on metallicity. 
An age/mass difference would be another possible
cause (possibly operating more strongly among the longer period
CephIIs like $\kappa$ Pav than among the shorter period one).   

If one adopts the results from the three
pulsation parallaxes, an LMC modulus of $18.55 \pm 0.15$ is
implied, neglecting any metallicity effect on CephII 
luminosities. 
This agrees with the RR Lyrae result given above which implies
a modulus of $18.55 \pm 0.12$. Neither of these values are
significantly different from the classical
Cepheid result ($18.39 \pm 0.05$).
However, the smaller distance for $\kappa$ Pav indicated by the  
revised Hipparcos result and the discussion of section 5.1, suggests
that, for the present, the results for this star should 
be viewed with some caution.
Additional pulsation parallaxes of CephIIs with periods near 10 days
and/or an improved trigonometrical parallax of $\kappa$ Pav would no doubt
throw more light on this problem.
\subsection{A Type II Cepheid distance scale}
 In section 5.2 we compared the Galactic CephII distance scale with that
implied by the Classical Cepheid scale (with metallicity corrections). In
this section we derive distance moduli for the LMC and for the Galactic
Centre, based directly on CephIIs. The two stars V553 Cen and SW Tau give a
mean zero-point, $c$ in eq. 4 of $-1.01 \pm 0.06$ where the standard error
comes from the standard errors of the two stars. If the pulsation parallax
result for $\kappa$ Pav is included the zero-point becomes $ c= -1.16 \pm
0.15$ where the standard error is based on the interagrement of the three
stars.

Matsunaga et al. (2006) list 2MASS, single-epoch,  $J,H,K_{s}$ 
photometry of LMC CephII stars with  known periods from Alcock
et al. (1998). There are 21 such stars with $\log P < 1.50$.
Longer period stars are not considered here as they may be
RV Tau stars. After correcting by $A_{K_{s}} = 0.02$ mag for absorption
these data were fitted to a line of the slope derived by Matsunaga et al.
(eq.4) viz:
\begin{equation}
K_{s}^{o} = -2.41 \log P + \gamma.
\end{equation}
We then find $\gamma = 17.31 \pm 0.08$ or if one somewhat discrepant
star is omitted $\gamma = 17.36 \pm 0.07$. With the values of $c$
in the previous paragraph these lead to the following estimates
of the LMC modulus: for 21 LMC stars, a modulus of $18.31 \pm 0.10$
from V553 Cen and SW Tau, or $18.47 \pm 0.17$ if we included 
$\kappa$ Pav.
Leaving out the discrepant LMC star we obtain for the two or three star
solutions, $18.37 \pm 0.09$ and $18.52 \pm 0.16$. Pending further
work on $\kappa$ Pav, the best value is probably $18.37 \pm 0.09$
but none of the values deviate significantly from the Classical
Cepheid value $18.39 \pm 0.05$.

Groenewegen et al. (2008) have recently estimated mean $K_{s}$ values
and periods for 39 CephIIs in the Galactic Bulge. After correction for
absorption they fit their data to an equation 
equivalent to eq. 8 above. Their result gives, $\gamma = 13.404 \pm 0.013$.
This together with the results for V553 Cen and SW Tau leads to
a modulus of the Galactic Centre of $14.42 \pm 0.06$ and to
a Galactic Centre distance of
$R_{0} = 7.64 \pm 0.21$ kpc. If we include $\kappa$ Pav we obtain
$14.56 \pm 0.15$ and $R_{0} = 8.18 \pm 0.56$ kpc. The first value, which
at present should probably be considered the preferred one, is close to 
that obtained by Eisenhauser et al. (2005) from the motion of a
star close to the central black-hole. With the suggested relativistic
correction of Zucker et al. (2006) this is, $R_{0} = 7.73 \pm 0.32$ kpc.
The value with $\kappa$ Pav included does not differ significantly
from this latter result.

\subsection{Optical period-luminosity relations}

The relations derived for CephIIs in the globular clusters NGC\,6441 and
NGC\,6388 (eqs. 5,6,7 above) at optical wavelengths, are quite narrow (see
Pritzl et al. 2003, fig. 8). On the other hand, plots of period-luminosity
diagrams in $B,V$ or $I$ for all known data for globular clusters and the
LMC (e.g. Pritzl et al. fig. 9) show very considerable scatter. Pritzl et
al. suggested that at least part of this scatter might be due to poor
photometry. This left open the question as to whether general PL relations
are as narrow as they found for their two clusters.  In Table 7 are the
deviations of our programme stars from eqs. 5,6,7. Table 7a gives the
results from the trigonometrical parallaxes and Table 7b those from the
pulsation parallaxes. In the case of the trigonometrical result for
$\kappa$ Pav the deviations are within the expected uncertainty (0.26)
whereas they are large for the pulsation parallax result which has a small
internal error (0.07).  As discussed in section 5.1 we prefer to leave a
solution of this matter to further work. The pulsation parallax results for
V553 Cen and SW Tau are of special interest since their formal uncertainties
are small (0.08). These two stars have deviations of opposite signs both
from the optical and infrared relations (Tables 7 and 6). The difference
between these two deviations thus gives an estimate of the lower limit of
the PL width at different wavelengths, independent of PL zero-point
considerations. These differences are: 0.77 mag at $B$, 0.51 at $V$, 0.20 at
$I$ and 0.07 at $K_{s}$. The results for VY Pyx, though of lower accuracy
agree with these results. This increase in the dispersion with decreasing
wavelength is, as in the case of classical Cepheids, naturally explained by
the existence of a finite instability strip.

The optical differences just 
quoted are significantly greater than the rms scatter about the PL
relations in NGC\,6441 and NGC\,6388 given by Pritzl et al. (2003) which are
0.10, 0.07 and 0.06 in $B, V, I$.  The possibility that the greater
optical differences estimated from V553 Cen and SW~Tau are due to the 
adoption of incorrect reddening corrections for these two stars
seems unlikely.  
%Any differences between the field an cluster stars cannot be explained
%in terms of general metallicity since all these objects are
%relatively metal-rich: NGC 6441 (-0.53), NGC 6388 (-0.60), SW Tau (+0.22),
%V553 Cen (+0.04), VY Pup (-0.44).  
The lower scatter in the case of the clusters is thus probably due
to the smaller range in the masses of the cluster variables compared with the
field.

The evolutionary state of the
metal-rich, short-period, CephIIs in the field has long
constituted something of a puzzle (see for instance section 4 of
Wallerstein 2002). As briefly summarized in section 1, the short
period CephII stars are thought to be moving through an instability
strip as they evolve from the blue HB towards the AGB. Old metal-rich
globular clusters have, in general, only stubby red HB and it is not
clear how stars of the ages and metallicities of these systems
could evolve into the CephII instability strip. NGC\,6441 and NGC\,6388
are well known as metal-rich systems which do have extended blue
HBs. There has been much discussion in the literature on the
cause of this anomaly in these and similar clusters. One possibility
is that the effect is due to enhanced helium abundance derived
from earlier generations of stars in the clusters (see for instance;
Lee et al. 2007, Caloi \& D'Antona 2006,  based on earlier work by
Rood  1973 and others). This seems unlikely to apply to field,
short-period, metal-rich, CephIIs. Thus either an alternative
explanation has to be found which will apply to both the field and cluster
stars, or, some 
other means will need to be found to move the field stars into
the instability strip.

%Table 5 lists
%the deviations of our absolute 
%magnitudes for VY Pyx 
%($\log P = 0.093$ and
%$\kappa$ Pav ($\log P = 0.959$   from eqs. 11-13 as well as from 
%the reddening-free
%relation, eq. 9.
%\begin{table}
%\centering
%\caption{Deviations from the relations for NGC6441 and NGC6388}
%\begin{tabular}{rrrr}
%\hline
%eq. & $\Delta$  & VY Pyx & $\kappa$ Pav\\
%12 & $\Delta M_{V}$ & +0.64 & -0.34 \\
%13 & $\Delta M_{B}$ & +0.89 & -0.25\\
%14 & $\Delta M_{I}$ &       & -0.10\\
%10 & $\Delta M_{V_{W}}$ &   & -0.15\\
%\hline
%\end{tabular}
%\end{table}
%TABLE 7 ******
\begin{table}
\centering
\caption{Deviation from optical relations.}\label{dev_opt}
\begin{tabular}{rrrr}
\hline
star & eq. 6 & eq. 5 & eq. 7\\
\hline
 & $\Delta M_{B}$ & $\Delta M_{V}$ & $\Delta M_{I}$\\
\hline
 & & (a) &\\
\hline
VY Pyx & +0.89 & +0.64 & \\
$\kappa$ Pav & -0.27 & -0.34 & -0.10 \\
\hline
&  & (b) &\\
\hline
$\kappa$ Pav & -0.77 & -0.77 & -0.60 \\
V553 Cen &  +0.56 & +0.26 & +0.09 \\
SW Tau & -0.21 & -0.25 & -0.11 \\
\hline
\end{tabular}

(a) Results using trigonometrical parallaxes.\\
(b) Results using pulsational parallaxes.\\ 
\end{table}

\section{Conclusions}
  Parallaxes of RR Lyrae variables from the revised Hipparcos catalogue
(van Leeuwen 2007) have been investigated. The parallax of
RR Lyrae itself obtained by combining the revised Hipparcos value with
an HST determination (Benedict et al. 2002) outweighs that of
all other members of the class. It yields $M_{K_{s}} = -0.64 \pm 0.11$
which is $0.16 \pm 0.12$ mag brighter than that implied by
observations of RR Lyrae variables in the LMC with a modulus of
$18.39 \pm 0.05$ derived from classical Cepheids (Benedict et al. 2007,
van Leeuwen et al. 2007 ). For 142 Hipparcos RR Lyrae variables 
mean $J,H,K_{s}$ based on phased-corrected 2MASS values are given.
These should be useful when discussing the proper motions and
radial velocities of the stars.
 Revised Hipparcos parallaxes for the CephIIs $\kappa$ Pav and VY Pyx are
given, and pulsation parallaxes for $\kappa$ Pav, V553 Cen and SW Tau
derived. Extensive new $J,H,K$ photometry of some of these stars and of some
classical Cepheids is tabulated. The latter data are used to establish 1.23
as the most appropriate ``$p$-factor" to use in the pulsational analysis of
Cepheids. The short-period, metal- and carbon-rich, disc population CephIIs
V553 Cen and SW Tau have pulsation-based absolute magnitudes of high
internal accuracy ($\pm 0.08$ mag). They fit closely (mean deviation 0.02
mag) the PL($K_{s}$) relation derived by Matsunaga et al. (2006) from
CephIIs in globular clusters and with a zero-point fixed by adopting an LMC
modulus of 18.39. The Hipparcos parallax of the short period star VY Pyx,
although it has higher uncertainty, agrees with this result. This suggests
that at least at short periods the CephIIs in the Galactic disc and in
Globular clusters fit the same PL($K_{s}$) relation rather closely. The
scatter of V553 Cen and SW Tau about the optical PL relations derived by
Pritzl et al. (2003) for the globular clusters NGC\,6388 and NGC\,6441 is
much greater than that about the Matsunaga PL($K_{s}$) relation, showing the
expected increase in PL widths with decreasing wavelength. This scatter
about the optical relations is also much greater than that of the CephIIs in
NGC\,6388/6441 themselves. Since the values of [Fe/H] are very similar for
V553 Cen and SW Tau this is unlikely to be due to a metallicity effect. It
presumably indicates a larger spread in masses for the short period CephIIs
in the general field than for those in the clusters.

The Hipparcos and pulsation parallaxes of the long-period star $\kappa$ Pav
differ by about 2$\sigma$. If the pulsation parallax is adopted, the value
of $M_{K_{s}}$ (which is of high internal accuracy,
$\sigma = 0.07$ mag) is more than $6\sigma$ from the Matsunaga relation with
a zero-point fixed by an LMC modulus of 18.39 and would suggest a
significant mass or metallicity effect at about this period ($\sim 10$
days). There are indications that this star may have a close companion. In
view of this, further work on the star and of others of similar period is
desirable before discussing in detail the implications for long-period
CephIIs. 

The results for V553 Cen and SW Tau together with published data on
CephIIs in the LMC and the Galactic Bulge lead to an LMC modulus of $18.37
\pm 0.09$ and to a distance to the Galactic centre of
$R_{0}=7.64 \pm 0.21$ kpc. Including the data for $\kappa$ Pav would
increase these estimates by $\sim 0.15$ mag.

\section*{Acknowledgments}

This publication makes use of data products from the Two Micron All Sky Survey,
which is a joint product of the University of Massachusetts and the Infrared
Processing and Analysis Center/California Institute of Technology, funded by
the National Aeronautics and Space Administration and the National Science
Foundation. We are grateful to the referee for their comments.

\setcounter{section}{1}
\renewcommand{\thesection}{A\arabic{section}}

\section*{APPENDIX A: Infrared photometry}
\renewcommand{\thetable}{A\arabic{table}}
\renewcommand{\thefigure}{A\arabic{figure}}
\setcounter{table}{0}
\setcounter{figure}{0}

Previously unpublished $JHK$ observations for classical Cepheids $\zeta$ Gem
and X Sgr, and for the CephIIs, V553 Cen, $\kappa$ Pav and SW Tau are
given in Table~A1. These data were obtained by CDL with 
the IRP Mk II
photometer and 0.75m telescope at the Sutherland observing station of the
South African Astronomical Observatory (SAAO), exactly as for the classical
Cepheid data given in Laney \& Stobie (1992). This single-channel device was
used with a 36 arcsec aperture and a chopping distance of 3 arcmin, and is
particularly suited to bright objects. The data are on the SAAO standard
system (Carter 1990), which was established with the same telescope,
photometer and filter set. Accuracy is typically 0.005-0.008 mag for bright
stars, including standardization. Similar data for l Car and $\beta$ Dor has
been taken from Laney \& Stobie (1992), while IR data for $\delta$ Cep on the
CIT system was taken from Barnes et al. (1997) and transformed to the Carter
system by the formulae given in Laney \& Stobie (1993). For convenience,
these data are given in Table A1 as well, with the phases and $V-K$
values calculated for the radius solutions used here.

\small
\onecolumn
\begin{longtable}{ccccccc}

\caption{Data for B-W solutions}\\ \hline

 \multicolumn{7}{c}{$\delta$ Cep}\\                              
 \multicolumn{7}{c}{Period 5.36624750d  Epoch 2448809.6246}\\ 
JD & Phase & $K$ & $H$ & $J$ & $V-K$ & $L$\\
2448429.980 & 0.2532 & 2.250 & 2.310 & 2.671 & 1.621 & \\
2448430.953 & 0.4346 & 2.275 & 2.346 & 2.761 & 1.817 & \\
2448431.985 & 0.6269 & 2.372 & 2.442 & 2.885 & 1.913 & \\
2448433.888 & 0.9815 & 2.301 & 2.349 & 2.612 & 1.219 & \\
2448434.969 & 0.1829 & 2.256 & 2.310 & 2.646 & 1.510 & \\
2448435.947 & 0.3652 & 2.273 & 2.338 & 2.732 & 1.747 & \\
2448436.958 & 0.5536 & 2.320 & 2.388 & 2.812 & 1.888 & \\
2448437.941 & 0.7368 & 2.444 & 2.521 & 2.939 & 1.900 & \\
2448438.912 & 0.9177 & 2.385 & 2.433 & 2.732 & 1.364 & \\
2448804.979 & 0.1343 & 2.254 & 2.309 & 2.632 & 1.439 & \\
2448805.964 & 0.3178 & 2.251 & 2.325 & 2.705 & 1.706 & \\
2448807.942 & 0.6864 & 2.411 & 2.484 & 2.917 & 1.921 & \\
2448808.930 & 0.8706 & 2.429 & 2.491 & 2.833 & 1.589 & \\
2448864.824 & 0.2864 & 2.231 & 2.296 & 2.668 & 1.685 & \\
2448865.799 & 0.4681 & 2.286 & 2.343 & 2.763 & 1.834 & \\
2448867.941 & 0.8673 & 2.434 & 2.494 & 2.849 & 1.601 & \\
2448870.747 & 0.3902 & 2.265 & 2.312 & 2.728 & 1.784 & \\
2448871.829 & 0.5918 & 2.349 & 2.419 & 2.847 & 1.901 & \\
2448872.920 & 0.7951 & 2.458 & 2.528 & 2.939 & 1.837 & \\
2448873.680 & 0.9367 & 2.343 & 2.393 & 2.669 & 1.309 & \\
2448873.976 & 0.9919 & 2.284 & 2.332 & 2.586 & 1.227 & \\
2449170.983 & 0.3391 & 2.250 & 2.309 & 2.705 & 1.736 & \\
2449172.912 & 0.6986 & 2.421 & 2.493 & 2.926 & 1.917 & \\
2449173.945 & 0.8911 & 2.401 & 2.462 & 2.783 & 1.500 & \\
2449174.952 & 0.0787 & 2.264 & 2.315 & 2.603 & 1.347 & \\
2449175.900 & 0.2554 & 2.250 & 2.311 & 2.673 & 1.624 & \\
2449176.984 & 0.4574 & 2.287 & 2.351 & 2.760 & 1.824 & \\ \hline

\multicolumn{7}{c}{X Sgr}\\                                   
\multicolumn{7}{c}{Period 7.0126750d Epoch 2449086.8197}\\ 
JD & Phase & $K$ & $H$ & $J$ & $V-K$ & $L$\\
2448846.455 & 0.7242 & 2.627 & 2.745 & 3.189 & 2.229 & \\
2448849.447 & 0.1509 & 2.451 & 2.541 & 2.918 & 1.928 & \\
2448850.461 & 0.2955 & 2.458 & 2.558 & 2.968 & 2.064 & \\
2448851.393 & 0.4284 & 2.487 & 2.591 & 3.025 & 2.188 & \\
2448852.429 & 0.5761 & 2.549 & 2.655 & 3.114 & 2.279 & \\
2449263.318 & 0.1685 & 2.444 & 2.541 & 2.923 & 1.956 & \\
2449291.232 & 0.1490 & 2.444 & 2.545 & 2.907 & 1.933 & \\
2449292.231 & 0.2914 & 2.452 & 2.547 & 2.961 & 2.068 & \\
2449534.463 & 0.8335 & 2.627 & 2.724 & 3.119 & 2.004 & \\
2449535.472 & 0.9773 & 2.509 & 2.603 & 2.927 & 1.756 & \\
2449537.453 & 0.2598 & 2.438 & 2.542 & 2.930 & 2.064 & \\
2449538.451 & 0.4021 & 2.473 & 2.572 & 2.996 & 2.160 & \\
2449598.329 & 0.9407 & 2.539 & 2.623 & 2.951 & 1.776 & \\
2449601.312 & 0.3661 & 2.469 & 2.559 & 2.982 & 2.111 & \\
2449859.677 & 0.2086 & 2.452 & 2.538 & 2.925 & 2.000 & \\
2449878.570 & 0.9028 & 2.588 & 2.671 & 3.015 & 1.823 & \\
2449889.482 & 0.4588 & 2.496 & 2.600 & 3.048 & 2.227 & \\
2449890.517 & 0.6064 & 2.561 & 2.667 & 3.128 & 2.277 & \\
2449941.461 & 0.8709 & 2.597 & 2.692 & 3.060 & 1.913 & \\
2449942.472 & 0.0151 & 2.478 & 2.571 & 2.900 & 1.782 & \\
2450142.667 & 0.5627 & 2.549 & 2.656 & 3.102 & 2.274 & \\
2450157.671 & 0.7023 & 2.618 & 2.717 & 3.165 & 2.246 & \\
2450160.679 & 0.1312 & 2.453 & 2.545 & 2.906 & 1.904 & \\
2450682.301 & 0.5139 & 2.510 & 2.612 & 3.060 & 2.280 & \\
2450683.405 & 0.6714 & 2.595 & 2.689 & 3.148 & 2.265 & \\
2450685.354 & 0.9493 & 2.516 & 2.610 & 2.936 & 1.783 & \\
2450912.682 & 0.3660 & 2.456 & 2.554 & 2.964 & 2.124 & \\
2450219.546 & 0.5256 & 2.533 & 2.630 & 3.088 & 2.268 & \\
2450221.553 & 0.8118 & 2.631 & 2.731 & 3.145 & 2.066 & \\
2450240.470 & 0.5093 & 2.505 & 2.622 & 3.069 & 2.282 & \\
2450244.557 & 0.0921 & 2.468 & 2.554 & 2.911 & 1.852 & \\
2449445.690 & 0.1745 & 2.444 & 2.542 & 2.906 & 1.965 & \\
2449590.389 & 0.8084 & 2.632 & 2.727 & 3.159 & 2.074 & \\
2449620.321 & 0.0767 & 2.476 & 2.558 & 2.910 & 1.830 & \\
2450262.466 & 0.6459 & 2.593 & 2.694 & 3.145 & 2.258 & \\
2450262.509 & 0.6520 & 2.578 & 2.692 & 3.140 & 2.275 & \\ \hline
\multicolumn{7}{c}{$\beta$ Dor}\\
\multicolumn{7}{c}{Period 9.842578  Epoch 2447913.2106}\\
JD & Phase & $K$ & $H$ & $J$ & $V-K$ & $L$\\
2447516.626 & 0.7072 & 2.073 & 2.131 & 2.537 & 1.742 & \\
2447517.537 & 0.7998 & 2.044 & 2.107 & 2.466 & 1.640 & \\
2447518.630 & 0.9108 & 1.987 & 2.041 & 2.401 & 1.627 & \\
2447520.269 & 0.0774 & 1.900 & 1.968 & 2.358 & 1.643 & \\
2447521.605 & 0.2131 & 1.856 & 1.922 & 2.348 & 1.825 & \\
2447522.287 & 0.2824 & 1.877 & 1.945 & 2.381 & 1.911 & \\
2447524.511 & 0.5084 & 1.994 & 2.070 & 2.531 & 2.087 & \\
2447525.362 & 0.5948 & 2.054 & 2.134 & 2.579 & 1.981 & \\
2447526.346 & 0.6948 & 2.057 & 2.136 & 2.531 & 1.772 & \\
2447528.533 & 0.9170 & 1.983 & 2.049 & 2.419 & 1.622 & \\
2447534.370 & 0.5100 & 1.995 & 2.076 & 2.562 & 2.086 & 1.894\\
2447567.352 & 0.8610 & 2.010 & 2.076 & 2.448 & 1.637 & \\
2447570.367 & 0.1673 & 1.869 & 1.929 & 2.323 & 1.760 & \\
2447604.320 & 0.6169 & 2.055 & 2.136 & 2.560 & 1.948 & \\
2447607.269 & 0.9165 & 1.982 & 2.044 & 2.415 & 1.623 & \\
2447642.251 & 0.4707 & 1.979 & 2.050 & 2.528 & 2.072 & \\
2447643.229 & 0.5700 & 2.045 & 2.124 & 2.589 & 2.008 & \\
2447644.221 & 0.6708 & 2.058 & 2.132 & 2.531 & 1.811 & \\
2447645.203 & 0.7706 & 2.051 & 2.107 & 2.488 & 1.680 & \\
2447646.194 & 0.8713 & 1.998 & 2.070 & 2.450 & 1.646 & \\
2447647.249 & 0.9785 & 1.918 & 1.988 & 2.330 & 1.535 & \\
2447660.204 & 0.2947 & 1.883 & 1.948 & 2.380 & 1.924 & 1.790\\
2447670.194 & 0.3097 & 1.892 & 1.955 & 2.408 & 1.935 & \\
2447675.187 & 0.8169 & 2.023 & 2.088 & 2.459 & 1.643 & \\
2447676.193 & 0.9192 & 1.982 & 2.048 & 2.422 & 1.619 & \\
2447713.698 & 0.7296 & 2.054 & 2.137 & 2.514 & 1.739 & \\
2447714.684 & 0.8298 & 2.009 & 2.085 & 2.455 & 1.649 & \\
2447715.727 & 0.9358 & 1.972 & 2.037 & 2.371 & 1.592 & \\
2447716.716 & 0.0363 & 1.902 & 1.980 & 2.316 & 1.577 & \\
2447719.719 & 0.3414 & 1.891 & 1.972 & 2.420 & 1.982 & \\
2447727.698 & 0.1520 & 1.875 & 1.940 & 2.330 & 1.731 & \\
2447731.687 & 0.5573 & 2.041 & 2.113 & 2.589 & 2.019 & \\
2447742.689 & 0.6751 & 2.062 & 2.135 & 2.538 & 1.798 & \\
2447744.673 & 0.8767 & 2.003 & 2.067 & 2.445 & 1.639 & 1.897\\
2447759.672 & 0.4006 & 1.923 & 2.001 & 2.473 & 2.047 & 1.823\\
2447769.689 & 0.4183 & 1.941 & 2.009 & 2.494 & 2.048 & \\
2447803.622 & 0.8659 & 2.001 & 2.071 & 2.436 & 1.645 & \\
2447811.633 & 0.6798 & 2.058 & 2.134 & 2.533 & 1.793 & \\
2447815.547 & 0.0774 & 1.904 & 1.962 & 2.343 & 1.639 & 1.800\\
2447816.620 & 0.1865 & 1.868 & 1.939 & 2.347 & 1.785 & 1.770\\
2447821.586 & 0.6910 & 2.061 & 2.129 & 2.528 & 1.772 & 1.954\\
2447823.528 & 0.8883 & 1.996 & 2.057 & 2.443 & 1.641 & 1.903\\ \hline
\multicolumn{7}{c}{$\zeta$ Gem}\\
\multicolumn{7}{c}{Period 10.14992  Epoch 2450180.19683}\\
JD & Phase & $K$ & $H$ & $J$ & $V-K$ & $L$\\
2448317.332 & 0.4651 & 2.142 & 2.228 & 2.697 & 2.029 & \\
2448318.339 & 0.5643 & 2.205 & 2.278 & 2.730 & 1.944 & \\
2448320.323 & 0.7598 & 2.200 & 2.266 & 2.655 & 1.689 & \\
2449789.295 & 0.4872 & 2.157 & 2.238 & 2.703 & 2.022 & \\
2449790.260 & 0.5823 & 2.209 & 2.292 & 2.724 & 1.919 & \\
2450079.492 & 0.0783 & 2.076 & 2.149 & 2.523 & 1.653 & \\
2450082.476 & 0.3723 & 2.103 & 2.185 & 2.648 & 1.982 & \\
2450147.267 & 0.7557 & 2.192 & 2.268 & 2.656 & 1.701 & \\
2450149.304 & 0.9563 & 2.118 & 2.171 & 2.541 & 1.589 & \\
2450150.279 & 0.0524 & 2.085 & 2.140 & 2.522 & 1.626 & \\
2450151.269 & 0.1499 & 2.057 & 2.121 & 2.528 & 1.742 & \\
2450152.253 & 0.2469 & 2.054 & 2.120 & 2.551 & 1.867 & \\
2450155.274 & 0.5445 & 2.192 & 2.278 & 2.732 & 1.974 & \\
2450156.300 & 0.6456 & 2.220 & 2.287 & 2.697 & 1.815 & \\
2450157.270 & 0.7412 & 2.207 & 2.277 & 2.659 & 1.700 & \\
2450158.257 & 0.8384 & 2.163 & 2.225 & 2.597 & 1.651 & \\
2450159.274 & 0.9386 & 2.127 & 2.183 & 2.551 & 1.591 & \\
2450160.283 & 0.0380 & 2.080 & 2.142 & 2.509 & 1.623 & \\
2450161.268 & 0.1351 & 2.049 & 2.115 & 2.511 & 1.734 & \\
2450471.427 & 0.6929 & 2.208 & 2.282 & 2.680 & 1.757 & \\
2450472.418 & 0.7905 & 2.192 & 2.256 & 2.635 & 1.668 & \\
2450473.400 & 0.8872 & 2.140 & 2.212 & 2.573 & 1.624 & \\
2450474.396 & 0.9854 & 2.096 & 2.154 & 2.517 & 1.600 & \\
2450475.397 & 0.0840 & 2.058 & 2.144 & 2.509 & 1.675 & \\
2450476.386 & 0.1814 & 2.050 & 2.121 & 2.532 & 1.787 & \\
2450824.411 & 0.4699 & 2.165 & 2.236 & 2.705 & 2.009 & \\
2450827.407 & 0.7651 & 2.192 & 2.260 & 2.637 & 1.692 & \\
2450886.262 & 0.5636 & 2.196 & 2.284 & 2.722 & 1.953 & \\
2451155.527 & 0.0924 & 2.065 & 2.127 & 2.522 & 1.676 & \\
2451177.479 & 0.2552 & 2.072 & 2.141 & 2.584 & 1.860 & \\ \hline
\multicolumn{7}{c}{$\ell$ Car}\\ 
\multicolumn{7}{c}{Period 35.543270  Epoch 2446104.2086}\\
JD & Phase & $K$ & $H$ & $J$ & $V-K$ & $L$\\
2446575.359 & 0.2557 & 0.973 & 1.088 & 1.656 & 2.684 & \\
2446576.304 & 0.2823 & 0.971 & 1.088 & 1.659 & 2.719 & \\
2446597.218 & 0.8707 & 1.286 & 1.383 & 1.904 & 2.467 & 1.119\\
2446601.246 & 0.9840 & 1.129 & 1.242 & 1.694 & 2.225 & \\
2446603.261 & 0.0407 & 1.069 & 1.185 & 1.662 & 2.300 & \\
2446607.256 & 0.1531 & 1.012 & 1.118 & 1.648 & 2.509 & \\
2446609.251 & 0.2092 & 0.975 & 1.081 & 1.635 & 2.624 & \\
2446610.264 & 0.2377 & 0.990 & 1.097 & 1.642 & 2.645 & \\
2446611.230 & 0.2649 & 0.978 & 1.099 & 1.662 & 2.690 & \\
2446740.634 & 0.9057 & 1.220 & 1.341 & 1.812 & 2.345 & \\
2446741.617 & 0.9333 & 1.183 & 1.289 & 1.755 & 2.265 & \\
2446758.567 & 0.4102 & 1.007 & 1.133 & 1.736 & 2.867 & 0.843\\
2446782.560 & 0.0852 & 1.033 & 1.157 & 1.652 & 2.387 & 0.924\\
2446803.518 & 0.6749 & 1.150 & 1.280 & 1.879 & 2.917 & 0.980\\
2446834.413 & 0.5441 & 1.077 & 1.202 & 1.828 & 2.954 & 0.903\\
2446862.546 & 0.3356 & 0.985 & 1.099 & 1.682 & 2.787 & \\
2446863.548 & 0.3638 & 0.983 & 1.108 & 1.702 & 2.834 & \\
2446880.261 & 0.8340 & 1.274 & 1.396 & 1.938 & 2.651 & \\
2446881.505 & 0.8690 & 1.275 & 1.398 & 1.899 & 2.487 & \\
2446886.494 & 0.0094 & 1.098 & 1.210 & 1.664 & 2.253 & \\
2446898.356 & 0.3431 & 0.978 & 1.086 & 1.675 & 2.806 & 0.817\\
2446899.341 & 0.3708 & 0.974 & 1.099 & 1.698 & 2.854 & 0.826\\
2446914.315 & 0.7921 & 1.267 & 1.394 & 1.964 & 2.756 & 1.126\\
2446915.396 & 0.8225 & 1.274 & 1.403 & 1.966 & 2.689 & 1.129\\
2446967.254 & 0.2815 & 0.946 & 1.071 & 1.637 & 2.743 & 0.830\\
2446972.286 & 0.4231 & 1.010 & 1.129 & 1.726 & 2.876 & \\
2446978.260 & 0.5912 & 1.110 & 1.243 & 1.863 & 2.944 & \\
2446981.247 & 0.6752 & 1.192 & 1.318 & 1.907 & 2.874 & \\
2446982.239 & 0.7031 & 1.207 & 1.334 & 1.932 & 2.856 & \\
2446983.240 & 0.7313 & 1.229 & 1.352 & 1.936 & 2.826 & \\
2446985.207 & 0.7866 & 1.282 & 1.398 & 1.961 & 2.748 & 1.131\\ \hline
\multicolumn{7}{c}{$\kappa$ Pav}\\
\multicolumn{7}{c}{Period 9.0814  Epoch 2446684.0691}\\
JD & Phase & $K$ & $H$ & $J$ & $V-K$ & $L$\\
2445928.495 & 0.7998 & 3.039 & 3.102 & 3.396 & 1.399 & \\
2445929.486 & 0.9089 & 2.816 & 2.886 & 3.149 & 1.177 & 2.778\\
2445953.440 & 0.5466 & 2.861 & 2.945 & 3.355 & 1.896 & \\
2446329.287 & 0.9331 & 2.799 & 2.856 & 3.105 & 1.154 & \\
2446345.338 & 0.7006 & 3.041 & 3.110 & 3.465 & 1.594 & \\
2446652.532 & 0.5273 & 2.859 & 2.934 & 3.354 & 1.893 & \\
2446675.477 & 0.0539 & 2.673 & 2.733 & 3.013 & 1.275 & \\
2446676.465 & 0.1627 & 2.646 & 2.715 & 3.045 & 1.518 & \\
2446680.420 & 0.5982 & 2.924 & 3.003 & 3.405 & 1.821 & \\
2446682.439 & 0.8205 & 3.012 & 3.067 & 3.353 & 1.330 & \\
2446684.440 & 0.0408 & 2.680 & 2.750 & 3.033 & 1.252 & \\
2446686.435 & 0.2605 & 2.654 & 2.732 & 3.111 & 1.740 & \\
2446694.370 & 0.1343 & 2.639 & 2.706 & 3.026 & 1.463 & 2.584\\
2446703.267 & 0.1140 & 2.667 & 2.730 & 3.051 & 1.391 & 2.612\\
2446739.269 & 0.0783 & 2.656 & 2.725 & 3.001 & 1.331 & \\
2446740.278 & 0.1895 & 2.645 & 2.716 & 3.058 & 1.580 & \\
2446741.275 & 0.2992 & 2.669 & 2.745 & 3.136 & 1.801 & \\
2446744.251 & 0.6269 & 2.952 & 3.019 & 3.420 & 1.763 & 2.868\\
2446748.245 & 0.0667 & 2.666 & 2.721 & 3.005 & 1.301 & \\
2446970.642 & 0.5560 & 2.875 & 2.953 & 3.366 & 1.883 & \\
2447029.481 & 0.0351 & 2.695 & 2.757 & 3.037 & 1.232 & \\
2447078.344 & 0.4157 & 2.737 & 2.818 & 3.238 & 1.942 & \\
2447646.640 & 0.9937 & 2.761 & 2.814 & 3.063 & 1.151 & \\
2447713.603 & 0.3673 & 2.729 & 2.802 & 3.202 & 1.863 & \\
2447715.598 & 0.5870 & 2.924 & 3.001 & 3.411 & 1.828 & \\ \hline
\multicolumn{7}{c}{V553 Cen}\\
\multicolumn{7}{c}{Period 2.060464  Epoch 2448437.11540}\\
JD & Phase & $K$ & $H$ & $J$ & $V-K$ & $L$\\
2446688.236 & 0.2206 & 6.755 & 6.851 & 7.180 & 1.631 & \\ 
2446864.611 & 0.8203 & 6.974 & 7.053 & 7.357 & 1.452 & \\ 
2446868.439 & 0.6781 & 6.992 & 7.083 & 7.408 & 1.595 & \\ 
2446881.644 & 0.0868 & 6.789 & 6.860 & 7.153 & 1.470 & \\ 
2446882.647 & 0.5736 & 6.973 & 7.073 & 7.431 & 1.725 & \\ 
2446886.628 & 0.5057 & 6.919 & 7.013 & 7.386 & 1.815 & \\ 
2446888.626 & 0.4754 & 6.900 & 6.991 & 7.367 & 1.814 & \\ 
2446890.616 & 0.4412 & 6.874 & 6.973 & 7.350 & 1.801 & \\ 
2446892.610 & 0.4089 & 6.852 & 6.939 & 7.309 & 1.782 & \\ 
2446978.388 & 0.0394 & 6.809 & 6.887 & 7.171 & 1.422 & \\ 
2446980.413 & 0.0222 & 6.819 & 6.893 & 7.186 & 1.407 & \\ 
2446981.365 & 0.4842 & 6.905 & 6.998 & 7.377 & 1.817 & \\ 
2446982.367 & 0.9705 & 6.862 & 6.942 & 7.223 & 1.365 & \\ 
2447029.269 & 0.7333 & 7.004 & 7.094 & 7.414 & 1.562 & \\ 
2447252.552 & 0.0987 & 6.784 & 6.858 & 7.155 & 1.485 & \\ 
2447602.594 & 0.9838 & 6.853 & 6.929 & 7.215 & 1.372 & \\ 
2447642.484 & 0.3435 & 6.810 & 6.906 & 7.260 & 1.742 & \\ 
2447643.521 & 0.8468 & 6.957 & 7.034 & 7.328 & 1.409 & \\ 
2447645.485 & 0.7999 & 6.990 & 7.064 & 7.365 & 1.481 & \\ 
2447646.482 & 0.2838 & 6.791 & 6.871 & 7.220 & 1.676 & \\ 
2447647.535 & 0.7949 & 6.989 & 7.067 & 7.371 & 1.493 & \\ 
2447674.414 & 0.8400 & 6.929 & 7.021 & 7.323 & 1.452 & \\ 
2447675.431 & 0.3336 & 6.795 & 6.881 & 7.239 & 1.743 & \\ 
2447676.431 & 0.8189 & 6.976 & 7.058 & 7.351 & 1.453 & \\ 
2447714.362 & 0.2278 & 6.775 & 6.861 & 7.196 & 1.619 & \\ 
2447716.363 & 0.1990 & 6.766 & 6.849 & 7.180 & 1.597 & \\ 
2447717.367 & 0.6863 & 7.005 & 7.083 & 7.411 & 1.579 & \\ 
2447771.267 & 0.8454 & 6.973 & 7.029 & 7.332 & 1.396 & \\ \hline 
\multicolumn{7}{c}{SW Tau}\\
\multicolumn{7}{c}{Period 1.583565d  Epoch 2445013.2696}\\
JD & Phase & $K$ & $H$ & $J$ & $V-K$ & $L$\\
2445950.650 & 0.9431 & 7.903 & 7.997 & 8.238 & 1.462 & \\
2445953.656 & 0.8414 & 7.979 & 8.065 & 8.286 & 1.414 & \\
2445954.643 & 0.4646 & 7.978 & 8.102 & 8.460 & 2.006 & \\
2446023.497 & 0.9450 & 7.897 & 7.988 & 8.239 & 1.467 & \\
2446024.407 & 0.5197 & 8.015 & 8.120 & 8.500 & 2.028 & \\
2446069.412 & 0.9397 & 7.896 & 7.990 & 8.245 & 1.472 & \\
2446073.364 & 0.4354 & 7.982 & 8.056 & 8.424 & 1.960 & \\
2446075.340 & 0.6832 & 8.173 & 8.262 & 8.588 & 1.892 & \\
2446326.631 & 0.3701 & 7.921 & 8.028 & 8.392 & 1.930 & \\
2446326.654 & 0.3846 & 7.919 & 8.020 & 8.382 & 1.951 & \\
2446334.623 & 0.4169 & 7.925 & 8.018 & 8.360 & 1.990 & \\
2446335.654 & 0.0680 & 7.857 & 7.945 & 8.200 & 1.553 & \\
2446338.664 & 0.9687 & 7.885 & 7.983 & 8.216 & 1.466 & \\
2446345.613 & 0.3569 & 7.920 & 8.012 & 8.364 & 1.915 & \\
2446363.537 & 0.6757 & 8.154 & 8.257 & 8.592 & 1.913 & \\
2446427.296 & 0.9387 & 7.919 & 7.986 & 8.224 & 1.449 & \\
2446664.636 & 0.8157 & 8.010 & 8.095 & 8.359 & 1.439 & \\
2446676.646 & 0.3998 & 7.933 & 8.027 & 8.384 & 1.957 & \\
2446677.645 & 0.0307 & 7.878 & 7.939 & 8.195 & 1.478 & \\
2446682.640 & 0.1850 & 7.905 & 7.991 & 8.319 & 1.768 & \\
2446686.649 & 0.7166 & 8.176 & 8.266 & 8.602 & 1.857 & \\
2446693.609 & 0.1117 & 7.890 & 7.984 & 8.283 & 1.614 & \\
2446702.618 & 0.8008 & 8.061 & 8.117 & 8.380 & 1.462 & \\
2446739.479 & 0.0780 & 7.878 & 7.968 & 8.237 & 1.552 & \\
2446740.568 & 0.7657 & 8.100 & 8.196 & 8.492 & 1.679 & \\
2446741.614 & 0.4262 & 7.937 & 8.063 & 8.411 & 1.991 & \\
2446744.481 & 0.2367 & 7.903 & 8.005 & 8.325 & 1.825 & \\
2446745.469 & 0.8606 & 7.959 & 8.045 & 8.295 & 1.431 & \\
2446746.536 & 0.5344 & 8.044 & 8.155 & 8.501 & 2.010 & \\
2446747.484 & 0.1331 & 7.881 & 7.978 & 8.267 & 1.673 & \\
2446748.506 & 0.7784 & 8.086 & 8.154 & 8.421 & 1.594 & \\
2446780.359 & 0.8932 & 7.942 & 8.057 & 8.283 & 1.454 & \\
2446783.391 & 0.8079 & 8.010 & 8.094 & 8.335 & 1.474 & \\
2446829.294 & 0.7950 & 8.042 & 8.127 & 8.383 & 1.517 & \\
2447023.659 & 0.5339 & 8.048 & 8.142 & 8.509 & 2.006 & \\
2447072.590 & 0.4331 & 7.951 & 8.053 & 8.425 & 1.987 & \\
2447077.644 & 0.6247 & 8.109 & 8.194 & 8.571 & 1.989 & \\
2447148.376 & 0.2910 & 7.911 & 8.007 & 8.333 & 1.853 & \\
2447431.631 & 0.1627 & 7.904 & 8.010 & 8.287 & 1.721 & \\
2447211.649 & 0.2470 & 7.891 & 7.996 & 8.346 & 1.843 & \\
2447212.640 & 0.8728 & 7.954 & 8.043 & 8.267 & 1.440 & \\
2447212.644 & 0.8754 & 7.948 & 8.034 & 8.273 & 1.447 & \\
2447212.661 & 0.8861 & 7.950 & 8.027 & 8.271 & 1.446 & \\
2447212.664 & 0.8880 & 7.946 & 8.034 & 8.273 & 1.450 & \\
2447212.706 & 0.9145 & 7.925 & 8.015 & 8.257 & 1.461 & \\
2447212.709 & 0.9164 & 7.919 & 8.015 & 8.253 & 1.466 & \\
2447219.617 & 0.2787 & 7.882 & 7.988 & 8.344 & 1.871 & \\
2447219.621 & 0.2812 & 7.884 & 7.990 & 8.333 & 1.871 & \\
2447219.678 & 0.3172 & 7.914 & 8.003 & 8.329 & 1.876 & \\
2447219.710 & 0.3374 & 7.903 & 8.007 & 8.349 & 1.910 & \\
2447220.655 & 0.9342 & 7.912 & 8.009 & 8.247 & 1.460 & \\
2447460.892 & 0.6406 & 8.125 & 8.228 & 8.570 & 1.962 & \\
2447460.974 & 0.6924 & 8.146 & 8.252 & 8.583 & 1.915 & \\
2447461.017 & 0.7196 & 8.171 & 8.274 & 8.601 & 1.855 & \\
2447462.792 & 0.8405 & 7.986 & 8.052 & 8.297 & 1.407 & \\
2447462.990 & 0.9655 & 7.902 & 7.986 & 8.221 & 1.450 & \\
2447465.796 & 0.7374 & 8.157 & 8.249 & 8.571 & 1.805 & \\
2447465.846 & 0.7690 & 8.101 & 8.193 & 8.485 & 1.652 & \\
2447465.921 & 0.8164 & 8.021 & 8.094 & 8.325 & 1.425 & \\
2447465.983 & 0.8555 & 7.955 & 8.044 & 8.281 & 1.434 & \\
2447466.023 & 0.8808 & 7.960 & 8.049 & 8.275 & 1.436 & \\
2447466.764 & 0.3487 & 7.922 & 8.019 & 8.356 & 1.904 & \\
2447466.828 & 0.3891 & 7.921 & 8.039 & 8.378 & 1.955 & \\
2447466.868 & 0.4144 & 7.947 & 8.050 & 8.401 & 1.964 & \\
2447466.906 & 0.4384 & 7.965 & 8.059 & 8.425 & 1.981 & \\
2447466.944 & 0.4624 & 7.987 & 8.073 & 8.434 & 1.994 & \\
2447466.985 & 0.4883 & 7.997 & 8.093 & 8.457 & 2.018 & \\
2447467.026 & 0.5142 & 8.028 & 8.134 & 8.486 & 2.011 & \\
2447468.838 & 0.6584 & 8.123 & 8.224 & 8.580 & 1.952 & \\
2447468.995 & 0.7576 & 8.110 & 8.208 & 8.531 & 1.729 & \\
2447469.019 & 0.7727 & 8.096 & 8.179 & 8.479 & 1.629 & \\
\end{longtable}

\renewcommand{\thesection}{B\arabic{section}}
\setcounter{section}{1}

\section*{Appendix B: The derivation of mean $JHK_{s}$ magnitudes for RR Lyrae stars 
with 2MASS magnitudes.}
\renewcommand{\thetable}{B\arabic{table}}
\renewcommand{\thefigure}{B\arabic{figure}}
\setcounter{table}{0}
\setcounter{figure}{0}

  The 2MASS Catalogue (Cutri et al. 2003) gives $JHK_{s}$ magnitudes for a single
  Julian Date. The derivation of the mean magnitudes (hereafter $<J>$,$<H>$,
  $<K_{s}>$) requires (a) ephemerides for each star that will give a phase for the
  2MASS data that is accurate to at least 0.1, (b) a visual amplitude
  ($\Delta V$) for the RR Lyrae star and (c) a standard light curve (or
  template) in each of
  $J$, $H$ \& $K_{s}$ which
  may be converted to the $J$,$H$ and $K_{s}$ light curves of the star in question
  by means of its $\Delta V$. Jones, Carney \& Fulbright (1996) gave templates
  of $K- <K>$ {\it vs} phase for a number of ranges of their $B$-amplitude for
type {\it ab} RR Lyrae stars; a single template was given for type {\it c}
variables. The method that we describe below covers $J$,$H$ and $K_{s}$ and gives
tables (rather than plots) from which the mean magnitudes can be computed.

\subsection{Ephemerides and visual amplitudes.}

   The 2MASS observations were made in the period 1997 to 2000. We therefore
  need to get a time of maximum light (JD(max)) for each variable that is
  as near to this epoch as possible. Fortunately a JD(max) for most of our
  variables can be found either in the ASAS catalogue (Pojmanski, 2002) which
  covers the sky south of declination +28$^{\circ}$ with epochs since 1997 or
  in the compilation by Wils et al. (2006) for epochs 1999 to 2000 for stars
  north of declination $-$38$^{\circ}$. In other cases, recent JD(max) are
  cited by Maintz (2005).  Periods were primarily taken from the ASAS catalogue
  (loc. cit.) or Maintz (loc. cit.).  The majority of the $\Delta V$s were taken
  from the catalogue of Nikolov, Buchantsova \&  Frolov (1984), the ASAS
  catalogue (loc. cit.) or Schmidt (1991). In some cases, the Hipparcos
  amplitude was multiplied by 0.874 to get $\Delta V$. This data allowed us to
  derive both the phase of the 2MASS data and $\Delta V$ for each variable.

\subsection{A standard RR Lyrae light curve for $J,H$ \&$K_{s}$}
 Jones, Carney \& Fulbright (1996) noted that the RR{\it ab} $K$ light curves
showed  small differences in their shapes that were a function of amplitude.
They therefore provided templates
of $K - <K>$ as a function of phase ($\phi$) for stars in five different
ranges of $B$-amplitude.
  These templates were derived from the $K$ light curves of
  field RR Lyrae stars that had been observed by several authors. We
  chose to produce a template of a single well observed RR Lyrae star (SW And)
of intermediate amplitude.  Excellent light curves in $J,H$ \& $K$ have been
  given for SW And by Barnes et al. (1992). These were based on observations
made in 1988; they also gave $BVRI$ data for the same year. Jones et al.
(1992) gave a partial $KBV$ light curve for SW And based on observations
made in 1987. In addition 31 unpublished observations in $H$ made by Kinman
between November 1987 and November 1989 were also available. All these
infrared observations were made using the Kitt Peak 1.3-m telescope
and are
shown in Fig.~B1 using the ephemeris:
\begin{equation}
    JD(max) = 24443067.6819  + 0.44226582 \times E
\end{equation}
The agreement between the three data sets shows that the light curve is stable.
The 2MASS observations were made twelve years later (JD = 24450739.8477) and the
phase (0.426) was determined from an ephemeris derived from the period given by
Maintz (2005) and the JD (max) given by Wils et al. (2006):
\begin{equation}
    JD(max) = 24451416.3203  + 0.442262 \times E
\end{equation}
The 2MASS observations (open squares) show good agreement with the light
curves given in Fig.~B1.  The intensity-weighted $<J>,<H>$ \& $<K>$ of SW
And are 8.780, 8.575 \& 8.575 respectively. The corrections to be applied to
the $J, H$ \& $K$
  magnitudes of this star as a function of phase to get the
  intensity-weighted mean magnitudes are given in Table B1.
  These corrections must be multiplied by a factor which
  takes into account the difference between the amplitude 
$\Delta V$ of SW And
  and that of the variable under consideration.

\begin{figure}
\includegraphics[width=8.5cm]{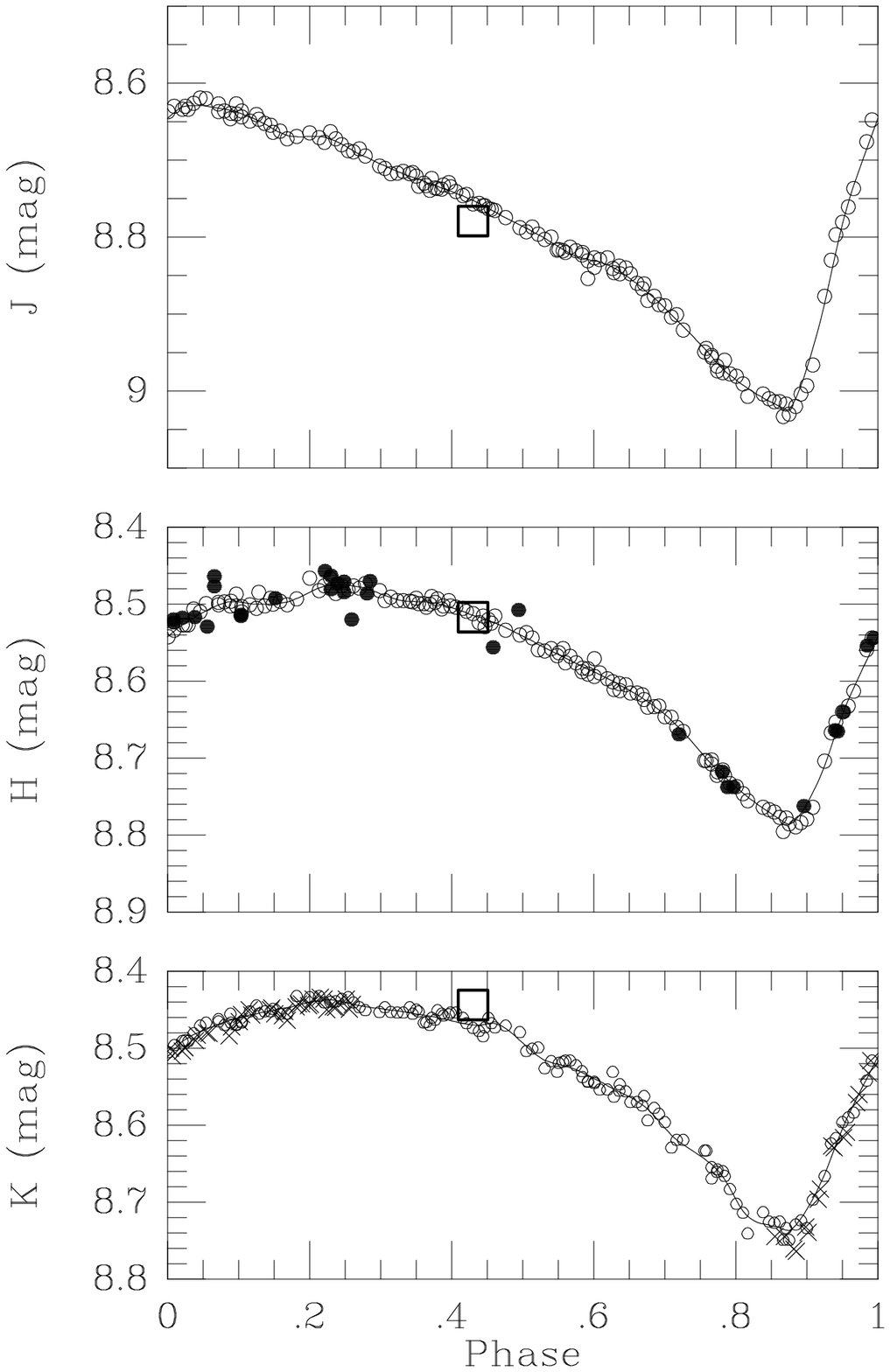}
 \caption{The $JHK$ light curves of SW And. The open circles are from
observations of Barnes et al. (1992), the crosses from those of Jones et al. 
(1992) and the filled circles are Kinman's unpublished observations. The
2MASS observations are shown as the large open squares.}
 \end{figure}

\subsection{The correction for the amplitude of the variable}
 A literature search for RR{\it ab} Lyrae stars with reliable infrared
light-curves
gave 11, 13 \& 27 with $J$,$H$ \& $K$ amplitudes respectively. These infrared
amplitudes are shown plotted against their corresponding $V$ amplitudes in
Fig.~B2 with the following linear fits:
\begin{equation}
\Delta J =  -0.015 + 0.450 \times \Delta V
\end{equation}
\begin{equation}
\Delta H  =  0.111 + 0.206 \times \Delta V
\end{equation}
\begin{equation}
 \Delta K =  0.176 + 0.125 \times \Delta V
\end{equation} 

\begin{center}
\onecolumn
\begin{longtable}{ccrcrcr}
\caption[Corrections to $JHK$ as function of phase.]
{Corrections to $JHK$ as function of phase.}\label{cor} \\
\hline
Phase & $J$ & \multicolumn{1}{c}{$\Delta J$} & $H$
& \multicolumn{1}{c}{$\Delta H$} & $K$ & \multicolumn{1}{c}{$\Delta K$} \\
\hline
\endfirsthead

\hline
Phase & $J$ & \multicolumn{1}{c}{$\Delta J$} & $H$
& \multicolumn{1}{c}{$\Delta H$} & $K$ & \multicolumn{1}{c}{$\Delta K$} \\
\hline
\endhead
%This is the footer for all pages except the last page of the table...
  \multicolumn{7}{l}{{Continued on Next Page\ldots}} \\
\endfoot    

%This is the footer for the last page of the table...
  \\ \hline 
\endlastfoot
  0.0000    &   8.6450   &    0.1350  &  8.5420   &    0.0330   &  8.5110    &   0.0210  \\
  0.0050    &   8.6415   &    0.1385  &  8.5383   &    0.0367   &  8.5061    &   0.0259   \\
  0.0100    &   8.6376   &    0.1424  &  8.5345   &    0.0405   &  8.5010    &   0.0310  \\
  0.0150    &   8.6339   &    0.1461  &  8.5308   &    0.0442   &  8.4960    &   0.0360  \\
  0.0200    &   8.6311   &    0.1489  &  8.5274   &    0.0476   &  8.4917    &   0.0403  \\
  0.0250    &   8.6299   &    0.1501  &  8.5245   &    0.0505   &  8.4884    &   0.0436  \\
  0.0300    &   8.6294   &    0.1506  &  8.5217   &    0.0533   &  8.4858    &   0.0462   \\
  0.0350    &   8.6291   &    0.1509  &  8.5189   &    0.0561   &  8.4834    &   0.0486  \\
  0.0400    &   8.6291   &    0.1509  &  8.5161   &    0.0589   &  8.4811    &   0.0509  \\
  0.0450    &   8.6292   &    0.1508  &  8.5134   &    0.0616   &  8.4789    &   0.0531  \\
  0.0500    &   8.6296   &    0.1504  &  8.5107   &    0.0643   &  8.4769    &   0.0551  \\
  0.0550    &   8.6300   &    0.1500  &  8.5081   &    0.0669   &  8.4750    &   0.0570  \\
  0.0600    &   8.6306   &    0.1494  &  8.5057   &    0.0693   &  8.4731    &   0.0589  \\
  0.0650    &   8.6313   &    0.1487  &  8.5035   &    0.0715   &  8.4714    &   0.0606  \\
  0.0700    &   8.6321   &    0.1479  &  8.5015   &    0.0735   &  8.4697    &   0.0623  \\
  0.0750    &   8.6330   &    0.1470  &  8.4997   &    0.0753   &  8.4680    &   0.0640   \\
  0.0800    &   8.6338   &    0.1462  &  8.4983   &    0.0767   &  8.4663    &   0.0657  \\
  0.0850    &   8.6348   &    0.1452  &  8.4972   &    0.0778   &  8.4647    &   0.0673  \\
  0.0900    &   8.6359   &    0.1441  &  8.4968   &    0.0782   &  8.4632    &   0.0688  \\
  0.0950    &   8.6372   &    0.1428  &  8.4969   &    0.0781   &  8.4619    &   0.0701   \\
  0.1000    &   8.6387   &    0.1413  &  8.4973   &    0.0777   &  8.4607    &   0.0713   \\
  0.1050    &   8.6403   &    0.1397  &  8.4979   &    0.0771   &  8.4595    &   0.0725  \\
  0.1100    &   8.6420   &    0.1380  &  8.4986   &    0.0764   &  8.4584    &   0.0736  \\
  0.1150    &   8.6438   &    0.1362  &  8.4993   &    0.0757   &  8.4573    &   0.0747  \\
  0.1200    &   8.6457   &    0.1343  &  8.4998   &    0.0752   &  8.4563    &   0.0757  \\
  0.1250    &   8.6476   &    0.1324  &  8.5000   &    0.0750   &  8.4552    &   0.0768  \\
  0.1300    &   8.6497   &    0.1303  &  8.5000   &    0.0750   &  8.4542    &   0.0778   \\
  0.1350    &   8.6521   &    0.1279  &  8.5000   &    0.0750   &  8.4532    &   0.0788   \\
  0.1400    &   8.6547   &    0.1253  &  8.4999   &    0.0751   &  8.4522    &   0.0798  \\
  0.1450    &   8.6575   &    0.1225  &  8.4998   &    0.0752   &  8.4513    &   0.0807  \\
  0.1500    &   8.6602   &    0.1198  &  8.4996   &    0.0754   &  8.4504    &   0.0816  \\
  0.1550    &   8.6627   &    0.1173  &  8.4993   &    0.0757   &  8.4495    &   0.0825   \\
  0.1600    &   8.6650   &    0.1150  &  8.4987   &    0.0763   &  8.4486    &   0.0834   \\
  0.1650    &   8.6669   &    0.1131  &  8.4979   &    0.0771   &  8.4477    &   0.0843  \\
  0.1700    &   8.6682   &    0.1118  &  8.4967   &    0.0783   &  8.4468    &   0.0852  \\
  0.1750    &   8.6691   &    0.1109  &  8.4952   &    0.0798   &  8.4458    &   0.0862   \\
  0.1800    &   8.6695   &    0.1105  &  8.4934   &    0.0816   &  8.4447    &   0.0873   \\
  0.1850    &   8.6697   &    0.1103  &  8.4913   &    0.0837   &  8.4435    &   0.0885  \\
  0.1900    &   8.6697   &    0.1103  &  8.4891   &    0.0859   &  8.4424    &   0.0896   \\
  0.1950    &   8.6695   &    0.1105  &  8.4868   &    0.0882   &  8.4412    &   0.0908  \\
  0.2000    &   8.6693   &    0.1107  &  8.4845   &    0.0905   &  8.4401    &   0.0919  \\
  0.2050    &   8.6692   &    0.1108  &  8.4822   &    0.0928   &  8.4391    &   0.0929  \\
  0.2100    &   8.6693   &    0.1107  &  8.4802   &    0.0948   &  8.4383    &   0.0937  \\
  0.2150    &   8.6695   &    0.1105  &  8.4784   &    0.0966   &  8.4376    &   0.0944   \\
  0.2200    &   8.6702   &    0.1098  &  8.4769   &    0.0981   &  8.4372    &   0.0948  \\
  0.2250    &   8.6713   &    0.1087  &  8.4758   &    0.0992   &  8.4370    &   0.0950   \\
  0.2300    &   8.6728   &    0.1072  &  8.4751   &    0.0999   &  8.4372    &   0.0948  \\
  0.2350    &   8.6747   &    0.1053  &  8.4747   &    0.1003   &  8.4377    &   0.0943   \\
  0.2400    &   8.6769   &    0.1031  &  8.4746   &    0.1004   &  8.4384    &   0.0936  \\
  0.2450    &   8.6793   &    0.1007  &  8.4746   &    0.1004   &  8.4394    &   0.0926  \\
  0.2500    &   8.6819   &    0.0981  &  8.4749   &    0.1001   &  8.4405    &   0.0915  \\
  0.2550    &   8.6845   &    0.0955  &  8.4754   &    0.0996   &  8.4416    &   0.0904   \\
  0.2600    &   8.6872   &    0.0928  &  8.4760   &    0.0990   &  8.4427    &   0.0893   \\
  0.2650    &   8.6897   &    0.0903  &  8.4768   &    0.0982   &  8.4438    &   0.0882  \\
  0.2700    &   8.6921   &    0.0879  &  8.4776   &    0.0974   &  8.4447    &   0.0873  \\
  0.2750    &   8.6943   &    0.0857  &  8.4786   &    0.0964   &  8.4454    &   0.0866  \\
  0.2800    &   8.6965   &    0.0835  &  8.4799   &    0.0951   &  8.4461    &   0.0859  \\
  0.2850    &   8.6987   &    0.0813  &  8.4813   &    0.0937   &  8.4468    &   0.0852  \\
  0.2900    &   8.7008   &    0.0792  &  8.4830   &    0.0920   &  8.4474    &   0.0846  \\
  0.2950    &   8.7030   &    0.0770  &  8.4848   &    0.0902   &  8.4480    &   0.0840  \\
  0.3000    &   8.7051   &    0.0749  &  8.4866   &    0.0884   &  8.4486    &   0.0834  \\
  0.3050    &   8.7072   &    0.0728  &  8.4884   &    0.0866   &  8.4492    &   0.0828  \\
  0.3100    &   8.7092   &    0.0708  &  8.4902   &    0.0848   &  8.4498    &   0.0822   \\
  0.3150    &   8.7113   &    0.0687  &  8.4919   &    0.0831   &  8.4504    &   0.0816  \\
  0.3200    &   8.7133   &    0.0667  &  8.4933   &    0.0817   &  8.4511    &   0.0809   \\
  0.3250    &   8.7152   &    0.0648  &  8.4946   &    0.0804   &  8.4517    &   0.0803  \\
  0.3300    &   8.7171   &    0.0629  &  8.4955   &    0.0795   &  8.4524    &   0.0796  \\
  0.3350    &   8.7189   &    0.0611  &  8.4961   &    0.0789   &  8.4531    &   0.0789   \\
  0.3400    &   8.7207   &    0.0593  &  8.4964   &    0.0786   &  8.4538    &   0.0782   \\
  0.3450    &   8.7224   &    0.0576  &  8.4965   &    0.0785   &  8.4545    &   0.0775  \\
  0.3500    &   8.7240   &    0.0560  &  8.4966   &    0.0784   &  8.4552    &   0.0768  \\
  0.3550    &   8.7256   &    0.0544  &  8.4966   &    0.0784   &  8.4559    &   0.0761   \\
  0.3600    &   8.7273   &    0.0527  &  8.4966   &    0.0784   &  8.4567    &   0.0753  \\
  0.3650    &   8.7289   &    0.0511  &  8.4969   &    0.0781   &  8.4575    &   0.0745  \\
  0.3700    &   8.7306   &    0.0494  &  8.4974   &    0.0776   &  8.4583    &   0.0737  \\
  0.3750    &   8.7324   &    0.0476  &  8.4982   &    0.0768   &  8.4592    &   0.0728  \\
  0.3800    &   8.7341   &    0.0459  &  8.4992   &    0.0758   &  8.4601    &   0.0719   \\
  0.3850    &   8.7358   &    0.0442  &  8.5004   &    0.0746   &  8.4612    &   0.0708   \\
  0.3900    &   8.7376   &    0.0424  &  8.5018   &    0.0732   &  8.4623    &   0.0697  \\
  0.3950    &   8.7393   &    0.0407  &  8.5032   &    0.0718   &  8.4634    &   0.0686   \\
  0.4000    &   8.7410   &    0.0390  &  8.5047   &    0.0703   &  8.4646    &   0.0674  \\
  0.4050    &   8.7428   &    0.0372  &  8.5063   &    0.0687   &  8.4657    &   0.0663  \\
  0.4100    &   8.7446   &    0.0354  &  8.5079   &    0.0671   &  8.4668    &   0.0652   \\
  0.4150    &   8.7465   &    0.0335  &  8.5095   &    0.0655   &  8.4678    &   0.0642   \\
  0.4200    &   8.7484   &    0.0316  &  8.5111   &    0.0639   &  8.4687    &   0.0633  \\
  0.4250    &   8.7504   &    0.0296  &  8.5127   &    0.0623   &  8.4695    &   0.0625  \\
  0.4300    &   8.7524   &    0.0276  &  8.5143   &    0.0607   &  8.4701    &   0.0619  \\
  0.4350    &   8.7546   &    0.0254  &  8.5158   &    0.0592   &  8.4702    &   0.0618   \\
  0.4400    &   8.7570   &    0.0230  &  8.5173   &    0.0577   &  8.4697    &   0.0623   \\
  0.4450    &   8.7594   &    0.0206  &  8.5189   &    0.0561   &  8.4689    &   0.0631  \\
  0.4500    &   8.7619   &    0.0181  &  8.5205   &    0.0545   &  8.4682    &   0.0638   \\
  0.4550    &   8.7645   &    0.0155  &  8.5221   &    0.0529   &  8.4678    &   0.0642   \\
  0.4600    &   8.7670   &    0.0130  &  8.5239   &    0.0511   &  8.4679    &   0.0641  \\
  0.4650    &   8.7695   &    0.0105  &  8.5258   &    0.0492   &  8.4689    &   0.0631  \\
  0.4700    &   8.7720   &    0.0080  &  8.5278   &    0.0472   &  8.4709    &   0.0611  \\
  0.4750    &   8.7744   &    0.0056  &  8.5300   &    0.0450   &  8.4737    &   0.0583  \\
  0.4800    &   8.7768   &    0.0032  &  8.5322   &    0.0428   &  8.4770    &   0.0550   \\
  0.4850    &   8.7791   &    0.0009  &  8.5345   &    0.0405   &  8.4808    &   0.0512  \\
  0.4900    &   8.7815   &   --0.0015  &  8.5368   &    0.0382   &  8.4848    &   0.0472  \\
  0.4950    &   8.7839   &   --0.0039  &  8.5392   &    0.0358   &  8.4891    &   0.0429  \\
  0.5000    &   8.7863   &   --0.0063  &  8.5415   &    0.0335   &  8.4934    &   0.0386  \\
  0.5050    &   8.7887   &   --0.0087  &  8.5440   &    0.0310   &  8.4977    &   0.0343   \\
  0.5100    &   8.7911   &   --0.0111  &  8.5464   &    0.0286   &  8.5019    &   0.0301  \\
  0.5150    &   8.7935   &   --0.0135  &  8.5488   &    0.0262   &  8.5058    &   0.0262   \\
  0.5200    &   8.7960   &   --0.0160  &  8.5512   &    0.0238   &  8.5093    &   0.0227  \\
  0.5250    &   8.7985   &   --0.0185  &  8.5535   &    0.0215   &  8.5124    &   0.0196   \\
  0.5300    &   8.8011   &   --0.0211  &  8.5559   &    0.0191   &  8.5150    &   0.0170  \\
  0.5350    &   8.8038   &   --0.0238  &  8.5582   &    0.0168   &  8.5170    &   0.0150   \\
  0.5400    &   8.8067   &   --0.0267  &  8.5606   &    0.0144   &  8.5186    &   0.0134   \\
  0.5450    &   8.8095   &   --0.0295  &  8.5629   &    0.0121   &  8.5199    &   0.0121   \\
  0.5500    &   8.8124   &   --0.0324  &  8.5653   &    0.0097   &  8.5210    &   0.0110   \\
  0.5550    &   8.8152   &   --0.0352  &  8.5676   &    0.0074   &  8.5221    &   0.0099  \\
  0.5600    &   8.8179   &   --0.0379  &  8.5700   &    0.0050   &  8.5232    &   0.0088  \\
  0.5650    &   8.8205   &   --0.0405  &  8.5723   &    0.0027   &  8.5244    &   0.0076  \\
  0.5700    &   8.8228   &   --0.0428  &  8.5746   &    0.0004   &  8.5259    &   0.0061  \\
  0.5750    &   8.8248   &   --0.0448  &  8.5769   &   --0.0019   &  8.5278    &   0.0042  \\
  0.5800    &   8.8264   &   --0.0464  &  8.5792   &   --0.0042   &  8.5298    &   0.0022   \\
  0.5850    &   8.8277   &   --0.0477  &  8.5814   &   --0.0064   &  8.5320    &   0.0000  \\
  0.5900    &   8.8287   &   --0.0487  &  8.5836   &   --0.0086   &  8.5343    &  --0.0023  \\
  0.5950    &   8.8297   &   --0.0497  &  8.5857   &   --0.0107   &  8.5366    &  --0.0046  \\
  0.6000    &   8.8306   &   --0.0506  &  8.5879   &   --0.0129   &  8.5390    &  --0.0070   \\
  0.6050    &   8.8316   &   --0.0516  &  8.5901   &   --0.0151   &  8.5415    &  --0.0095  \\
  0.6100    &   8.8328   &   --0.0528  &  8.5924   &   --0.0174   &  8.5440    &  --0.0120  \\
  0.6150    &   8.8343   &   --0.0543  &  8.5947   &   --0.0197   &  8.5465    &  --0.0145  \\
  0.6200    &   8.8361   &   --0.0561  &  8.5970   &   --0.0220   &  8.5490    &  --0.0170  \\
  0.6250    &   8.8384   &   --0.0584  &  8.5995   &   --0.0245   &  8.5515    &  --0.0195   \\
  0.6300    &   8.8410   &   --0.0610  &  8.6019   &   --0.0269   &  8.5538    &  --0.0218  \\
  0.6350    &   8.8438   &   --0.0638  &  8.6043   &   --0.0293   &  8.5561    &  --0.0241  \\
  0.6400    &   8.8468   &   --0.0668  &  8.6067   &   --0.0317   &  8.5583    &  --0.0263  \\
  0.6450    &   8.8499   &   --0.0699  &  8.6092   &   --0.0342   &  8.5606    &  --0.0286    \\
 0.6500     &  8.8533    &  --0.0733   & 8.6117    &  --0.0367    & 8.5630     & --0.0310   \\
 0.6550     &  8.8567    &  --0.0767   & 8.6143    &  --0.0393    & 8.5655     & --0.0335   \\
 0.6600     &  8.8602    &  --0.0802   & 8.6170    &  --0.0420    & 8.5683     & --0.0363    \\
 0.6650     &  8.8639    &  --0.0839   & 8.6198    &  --0.0448    & 8.5713     & --0.0393   \\
 0.6700     &  8.8675    &  --0.0875   & 8.6228    &  --0.0478    & 8.5746     & --0.0426   \\
 0.6750     &  8.8712    &  --0.0912   & 8.6259    &  --0.0509    & 8.5783     & --0.0463   \\
 0.6800     &  8.8752    &  --0.0952   & 8.6291    &  --0.0541    & 8.5827     & --0.0507   \\
 0.6850     &  8.8792    &  --0.0992   & 8.6324    &  --0.0574    & 8.5877     & --0.0557    \\
 0.6900     &  8.8834    &  --0.1034   & 8.6358    &  --0.0608    & 8.5929     & --0.0609   \\
 0.6950     &  8.8878    &  --0.1078   & 8.6394    &  --0.0644    & 8.5983     & --0.0663   \\
 0.7000     &  8.8922    &  --0.1122   & 8.6432    &  --0.0682    & 8.6037     & --0.0717   \\
 0.7050     &  8.8967    &  --0.1167   & 8.6472    &  --0.0722    & 8.6088     & --0.0768   \\
 0.7100     &  8.9013    &  --0.1213   & 8.6514    &  --0.0764    & 8.6135     & --0.0815   \\
 0.7150     &  8.9058    &  --0.1258   & 8.6558    &  --0.0808    & 8.6176     & --0.0856   \\
 0.7200     &  8.9104    &  --0.1304   & 8.6605    &  --0.0855    & 8.6213     & --0.0893   \\
 0.7250     &  8.9151    &  --0.1351   & 8.6654    &  --0.0904    & 8.6247     & --0.0927   \\
 0.7300     &  8.9199    &  --0.1399   & 8.6705    &  --0.0955    & 8.6279     & --0.0959    \\
 0.7350     &  8.9246    &  --0.1446   & 8.6757    &  --0.1007    & 8.6309     & --0.0989   \\
 0.7400     &  8.9295    &  --0.1495   & 8.6810    &  --0.1060    & 8.6339     & --0.1019   \\
 0.7450     &  8.9343    &  --0.1543   & 8.6863    &  --0.1113    & 8.6369     & --0.1049   \\
 0.7500     &  8.9392    &  --0.1592   & 8.6916    &  --0.1166    & 8.6399     & --0.1079    \\
 0.7550     &  8.9441    &  --0.1641   & 8.6969    &  --0.1219    & 8.6431     & --0.1111    \\
 0.7600     &  8.9490    &  --0.1690   & 8.7020    &  --0.1270    & 8.6466     & --0.1146   \\
 0.7650     &  8.9541    &  --0.1741   & 8.7070    &  --0.1320    & 8.6503     & --0.1183   \\
 0.7700     &  8.9597    &  --0.1797   & 8.7119    &  --0.1369    & 8.6543     & --0.1223   \\
 0.7750     &  8.9652    &  --0.1852   & 8.7167    &  --0.1417    & 8.6591     & --0.1271   \\
 0.7800     &  8.9701    &  --0.1901   & 8.7211    &  --0.1461    & 8.6647     & --0.1327   \\
 0.7850     &  8.9743    &  --0.1943   & 8.7254    &  --0.1504    & 8.6716     & --0.1396    \\
 0.7900     &  8.9782    &  --0.1982   & 8.7296    &  --0.1546    & 8.6795     & --0.1475    \\
 0.7950     &  8.9819    &  --0.2019   & 8.7337    &  --0.1587    & 8.6880     & --0.1560   \\
 0.8000     &  8.9855    &  --0.2055   & 8.7378    &  --0.1628    & 8.6964     & --0.1644   \\
 0.8050     &  8.9890    &  --0.2090   & 8.7418    &  --0.1668    & 8.7045     & --0.1725   \\
 0.8100     &  8.9922    &  --0.2122   & 8.7456    &  --0.1706    & 8.7115     & --0.1795    \\
 0.8150     &  8.9954    &  --0.2154   & 8.7493    &  --0.1743    & 8.7171     & --0.1851    \\
 0.8200     &  8.9984    &  --0.2184   & 8.7528    &  --0.1778    & 8.7211     & --0.1891   \\
 0.8250     &  9.0015    &  --0.2215   & 8.7564    &  --0.1814    & 8.7240     & --0.1920   \\
 0.8300     &  9.0044    &  --0.2244   & 8.7598    &  --0.1848    & 8.7262     & --0.1942    \\
 0.8350     &  9.0073    &  --0.2273   & 8.7631    &  --0.1881    & 8.7277     & --0.1957    \\
 0.8400     &  9.0100    &  --0.2300   & 8.7664    &  --0.1914    & 8.7288     & --0.1968   \\
 0.8450     &  9.0125    &  --0.2325   & 8.7695    &  --0.1945    & 8.7296     & --0.1976    \\
 0.8500     &  9.0149    &  --0.2349   & 8.7725    &  --0.1975    & 8.7304     & --0.1984   \\
 0.8550     &  9.0169    &  --0.2369   & 8.7754    &  --0.2004    & 8.7313     & --0.1993   \\
 0.8600     &  9.0189    &  --0.2389   & 8.7782    &  --0.2032    & 8.7326     & --0.2006   \\
 0.8650     &  9.0217    &  --0.2417   & 8.7817    &  --0.2067    & 8.7346     & --0.2026   \\
 0.8700     &  9.0240    &  --0.2440   & 8.7848    &  --0.2098    & 8.7364     & --0.2044    \\
 0.8750     &  9.0235    &  --0.2435   & 8.7860    &  --0.2110    & 8.7366     & --0.2046   \\
 0.8800     &  9.0180    &  --0.2380   & 8.7840    &  --0.2090    & 8.7340     & --0.2020    \\
 0.8850     &  9.0084    &  --0.2284   & 8.7793    &  --0.2043    & 8.7291     & --0.1971   \\
 0.8900     &  8.9971    &  --0.2171   & 8.7735    &  --0.1985    & 8.7233     & --0.1913    \\
 0.8950     &  8.9843    &  --0.2043   & 8.7668    &  --0.1918    & 8.7166     & --0.1846   \\
 0.9000     &  8.9702    &  --0.1902   & 8.7592    &  --0.1842    & 8.7093     & --0.1773   \\
 0.9050     &  8.9551    &  --0.1751   & 8.7508    &  --0.1758    & 8.7013     & --0.1693   \\
 0.9100     &  8.9392    &  --0.1592   & 8.7417    &  --0.1667    & 8.6927     & --0.1607    \\
 0.9150     &  8.9227    &  --0.1427   & 8.7320    &  --0.1570    & 8.6837     & --0.1517    \\
 0.9200     &  8.9055    &  --0.1255   & 8.7216    &  --0.1466    & 8.6742     & --0.1422   \\
 0.9250     &  8.8862    &  --0.1062   & 8.7098    &  --0.1348    & 8.6635     & --0.1315   \\
 0.9300     &  8.8656    &  --0.0856   & 8.6969    &  --0.1219    & 8.6520     & --0.1200   \\
 0.9350     &  8.8442    &  --0.0642   & 8.6834    &  --0.1084    & 8.6400     & --0.1080   \\
 0.9400     &  8.8228    &  --0.0428   & 8.6696    &  --0.0946    & 8.6279     & --0.0959   \\
 0.9450     &  8.8021    &  --0.0221   & 8.6562    &  --0.0812    & 8.6159     & --0.0839   \\
 0.9500     &  8.7829    &  --0.0029   & 8.6434    &  --0.0684    & 8.6045     & --0.0725   \\
 0.9550     &  8.7658    &   0.0142   & 8.6318    &  --0.0568    & 8.5940     & --0.0620   \\
 0.9600     &  8.7502    &   0.0298   & 8.6209    &  --0.0459    & 8.5841     & --0.0521   \\
 0.9650     &  8.7357    &   0.0443   & 8.6105    &  --0.0355    & 8.5745     & --0.0425    \\
 0.9700     &  8.7220    &   0.0580   & 8.6004    &  --0.0254    & 8.5652     & --0.0332   \\
 0.9750     &  8.7089    &   0.0711   & 8.5905    &  --0.0155    & 8.5560     & --0.0240    \\
 0.9800     &  8.6961    &   0.0839   & 8.5808    &  --0.0058    & 8.5470     & --0.0150   \\
 0.9850     &  8.6836    &   0.0964   & 8.5712    &   0.0038    & 8.5381     & --0.0061   \\
 0.9900     &  8.6710    &   0.1090   & 8.5616    &   0.0134    & 8.5292     &  0.0028    \\
 0.9950     &  8.6582    &   0.1218   & 8.5519    &   0.0231    & 8.5201     &  0.0119    \\
 1.0000     &  8.6450    &   0.1350   & 8.5420    &   0.0330    & 8.5110     &  0.0210   \\
\end{longtable}
\end{center}
\twocolumn

  The $J$, $H$ \& $K$ amplitudes of SW And are 0.395, 0.314 \& 0.300 respectively.
  The corrections in Table~B1  must therefore be multiplied by the
  following factors for a type {\it ab} variable with amplitude $\Delta V$:
\begin{equation}
  -0.038  + 1.139 \times \Delta V   \ \ \ \ \ \ \rm for\ $J$;
\end{equation}
\begin{equation}
   0.358  + 0.665 \times \Delta V  \ \ \ \  \ \rm for \ $H$;
\end{equation}
\begin{equation}
   0.594  + 0.417 \times \Delta V \ \ \ \  \ \rm for \ $K$.
\end{equation}

  These corrections must be added to the observed magnitudes to obtain the
  mean magnitudes. In the case of RR{\it c} variables (which have quite low
  amplitudes) the above corrections can also be applied for the $J$ magnitudes,
  while the $K$ - $<K>$ correction of Jones et al. (loc. cit.) can be applied
  to both the $H$ and $K$ magnitudes to get the mean magnitudes.
\begin{figure}
\includegraphics[width=8.5cm]{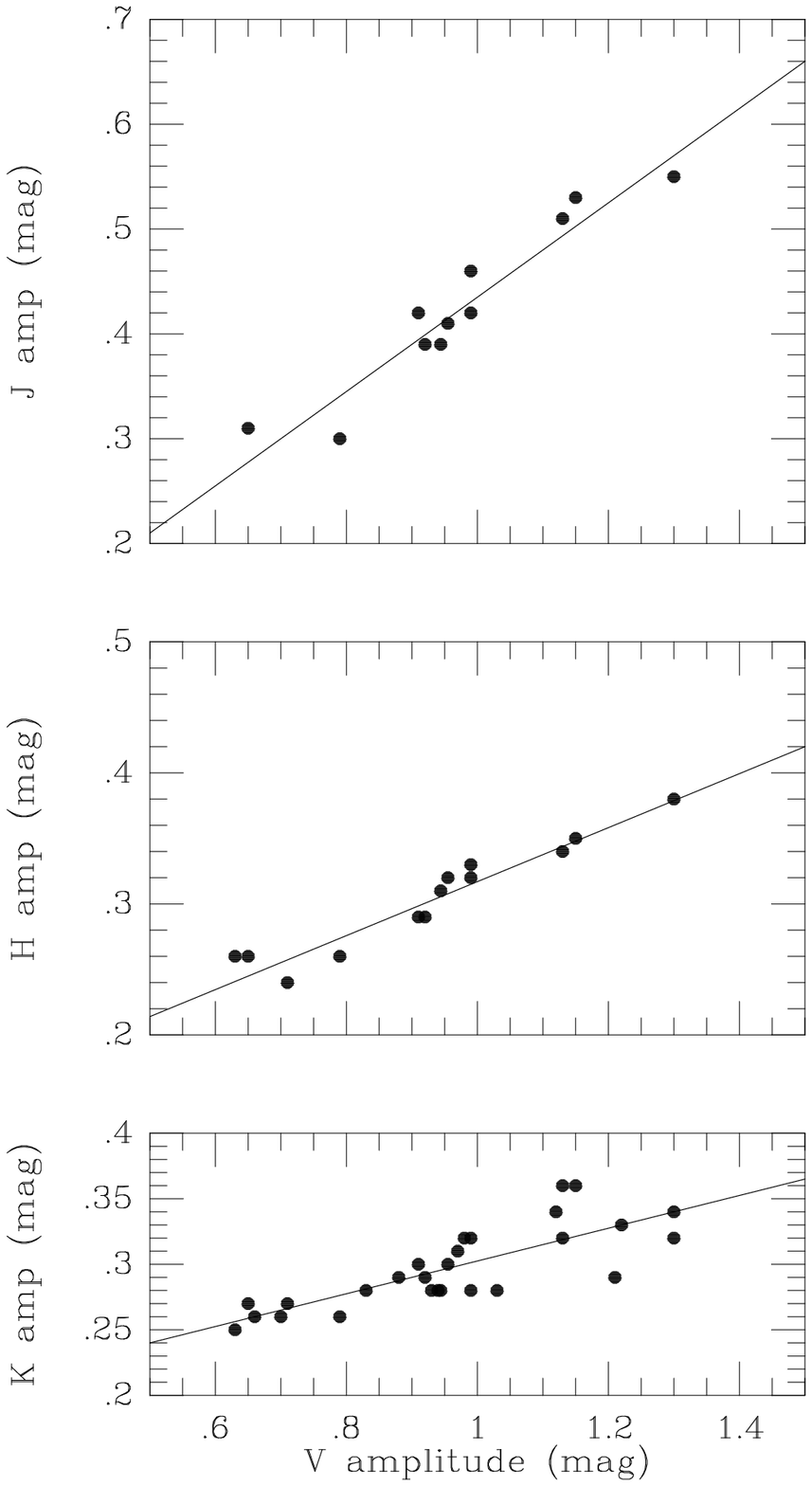}
\caption{The relation between the $J$, $H$ and $K$ amplitudes and their
$V$ amplitudes for RR Lyrae stars that have well determined light curves.}
\end{figure}

\subsection{The accuracy of these corrections}
 Table~\ref{t:comp} compares the mean magnitudes $<J>$,$<H>$ \& $<K_{s}>$ derived from 2MASS
data (Source 1) with those derived from the data of Fernley et al. (1993)
(Source 2) and from unpublished $H$ magnitudes of Kinman (Source 3). The largest
discrepancies are for RZ Cep which is multiperiodic and has a double-peaked
maximum (Cester \& Todoran 1976). The second observation of RR Lyrae
by Fernley et al. (1993) (indicated by an asterisk in Table~\ref{t:comp}) was taken near
maximum light. RR Lyrae shows a Blazhko effect of varying period so the large
discrepancy between this and the other two observations is not surprising.
  If we neglect these observations, the mean differences in
the sense (Fernley et al. {\it minus} 2MASS) are +0.006$\pm$0.013,
+0.008$\pm$0.015  \& +0.008$\pm$0.0015 mag for $<J>$,$<H>$ \& $<K_{s}>$
respectively.  The mean difference (Kinman {\it minus} 2MASS) is
$-$0.006$\pm$0.006 for $<H>$. These differences 
do not disagree with the small differences expected between observations
made using the standards of Elias et al. (1982, 1983)
as was the case of the Fernley et al. and the Kinman data
and those on the 2MASS system (Carpenter 2001). 
It must be remembered that errors of as much as 0.2 mag can occur near the
rising branch or with stars with varying light-curves and/or ephemerides.

\begin{table}
\caption{comparison of 2MASS mean magnitudes with those derived from other
  sources.}
\label{t:comp}
\begin{center}
\leavevmode
\begin{tabular}[h]{lcccc}
\noalign{\smallskip}
\hline
\noalign{\smallskip}
Star  &
$<J>$  &
$<H>$  &
$<K_{s}>$  &
Source  \\
          &  &  &  &        \\
  (1) & (2) & (3)& (4)& (5)  \\
\noalign{\smallskip}
\hline
\noalign{\smallskip}
           &          &           &           &             \\
  SW AND   & 8.807    & 8.578     & 8.506     &    (1)      \\
           &$\cdots$  & 8.573     & $\cdots$  &    (3)      \\
           &          &           &           &             \\
  TU UMA   & 8.907    & 8.728     & 8.654     &    (1)      \\
           &$\cdots$  & 8.714     & $\cdots$  &    (3)      \\
           &          &           &           &             \\
  BH PEG   & 9.345    & 9.085     & 9.041     &    (1)      \\
           & 9.345    & 9.065     & 9.025     &    (2)      \\
           & 9.395    & 9.055     & 9.009     &    (2)      \\
           &$\cdots$  & 9.106     & $\cdots$  &    (3)      \\
           &          &           &           &             \\
  RR LYR   & 6.739    & 6.511     & 6.462     &    (1)      \\
           & 6.780    & 6.530     & 6.490     &    (2)      \\
           & 6.930    & 6.670     & 6.650     &    (2)$^{\ast}$  \\
           &          &           &           &             \\
  SV ERI   & 8.947    & 8.703     & 8.636     &    (1)      \\
           & 8.915    & 8.672     & 8.658     &    (2)      \\
           & 8.934    & 8.700     & 8.615     &    (2)      \\
           &$\cdots$  & 8.682     & $\cdots$  &    (3)      \\
           &          &           &           &             \\
  RZ CEP   & 8.251    & 8.068     & 7.998     &    (1)      \\
           & 8.350    & 8.270     & 8.160     &    (2)      \\
           & 8.360    & 8.140     & 8.140     &    (2)      \\
           &          &           &           &             \\
  XZ CYG   & 8.914    & 8.751     & 8.682     &    (1)      \\
           & 8.970    & 8.790     & 8.770     &    (2)      \\
           & 8.890    & 8.820     & 8.680     &    (2)      \\
           &$\cdots$  & 8.730     & $\cdots$  &    (3)      \\
           &          &           &           &             \\
  DX DEL   & 9.001    & 8.741     & 8.682     &    (1)      \\
           &$\cdots$  & 8.736     & $\cdots$  &    (3)      \\
           &          &           &           &             \\
  X ARI    & 8.327    & 8.026     & 7.928     &    (1)      \\
           &$\cdots$  & 8.030     & $\cdots$  &    (3)      \\
           &          &           &           &             \\

\noalign{\smallskip}
\hline
\end{tabular}\end{center}\end{table}

\end{document}